\documentclass{article}

\usepackage{arxiv}

\usepackage[utf8]{inputenc} 
\usepackage[T1]{fontenc}    
\usepackage{hyperref}       
\usepackage{url}            
\usepackage{booktabs}       
\usepackage{amsfonts}       
\usepackage{nicefrac}       
\usepackage{microtype}      
\usepackage{lipsum}		
\usepackage{graphicx, subcaption}
\usepackage{natbib}
\usepackage{doi}
\usepackage{gensymb}
\usepackage{algorithm}
\usepackage{algpseudocode}
\usepackage{amsmath}
\usepackage{color,soul}
\usepackage{siunitx}
\usepackage{enumitem}
\newcommand{\vv}[1]{\textbf{#1}}

\makeatletter
\newcommand{\customlabel}[2]{%
   \protected@write \@auxout {}{\string \newlabel {#1}{{#2}{\thepage}{#2}{#1}{}} }%
   \hypertarget{#1}{}
}
\makeatother

\title{FourCastNet: A Global Data-driven High-resolution Weather Model using Adaptive Fourier Neural Operators}

\author{Jaideep Pathak \\
	NVIDIA Corporation\\
	Santa Clara, CA 95051 \\
	\And
	Shashank Subramanian \\
	Lawrence Berkeley \\National Laboratory\\
	Berkeley, CA 94720 \\
	\And
	Peter Harrington \\
	Lawrence Berkeley \\National Laboratory\\
	Berkeley, CA 94720 \\
	\And
	Sanjeev Raja \\
	University of Michigan\\
	Ann Arbor, MI 48109 \\
	\AND
	Ashesh Chattopadhyay \\
	Rice University \\
	Houston, TX 77005 \\
	\And
	Morteza Mardani \\
	NVIDIA Corporation\\
	Santa Clara, CA 95051 \\
	\And
	Thorsten Kurth \\
	NVIDIA Corporation\\
	Santa Clara, CA 95051 \\
	\AND
	David Hall \\
	NVIDIA Corporation\\
	Santa Clara, CA 95051 \\
	\And
	Zongyi Li \\
	California Institute of Technology \\
	Pasadena, CA 91125 \\
	NVIDIA Corporation\\
	Santa Clara, CA 95051
	\And
	Kamyar Azizzadenesheli \\
	Purdue University\\
	West Lafayette, IN 47907 \\
	\AND
	Pedram Hassanzadeh \\
	Rice University\\
	Houston, TX 77005 \\
	\And
	Karthik Kashinath \\
	NVIDIA Corporation\\
	Santa Clara, CA 95051 \\
	\And
	Animashree Anandkumar \\
	California Institute of Technology \\
	Pasadena, CA 91125 \\
	NVIDIA Corporation\\
	Santa Clara, CA 95051
}

\renewcommand{\shorttitle}{FourCastNet: A global data-driven high-resolution weather model}

\hypersetup{
pdftitle={High-resolution-global-weather-model-afno},
pdfkeywords={Numerical Weather Prediction, Deep Learning, Fourier Neural Operator, Transformer},
}

\begin{document}
\maketitle

\begin{abstract}
FourCastNet, short for \textit{Four}ier Fore\textit{Cast}ing Neural \textit{Net}work, is a global data-driven weather forecasting model that provides accurate short to medium-range global predictions at $0.25^{\circ}$ resolution. FourCastNet accurately forecasts high-resolution, fast-timescale variables such as the surface wind speed, precipitation, and atmospheric water vapor. It has important implications for planning wind energy resources, predicting extreme weather events such as tropical cyclones, extra-tropical cyclones, and atmospheric rivers. FourCastNet matches the forecasting accuracy of the ECMWF Integrated Forecasting System (IFS), a state-of-the-art Numerical Weather Prediction (NWP) model, at short lead times for large-scale variables, while outperforming IFS for small-scale variables, including precipitation. FourCastNet generates a week-long forecast in less than 2 seconds, orders of magnitude faster than IFS. The speed of FourCastNet enables the creation of rapid and inexpensive large-ensemble forecasts with thousands of ensemble-members for improving probabilistic forecasting. We discuss how data-driven deep learning models such as FourCastNet are a valuable addition to the meteorology toolkit to aid and augment NWP models.
\end{abstract}


\section{Introduction}\label{sec:introduction}

The beginnings of modern numerical weather prediction (NWP) can be traced to the 1920s. Now ubiquitous, they contribute to economic planning in key sectors such as transport, logistics, agriculture, and energy production. Accurate weather forecasts have saved countless human lives by providing advance notice of extreme events. The quality of weather forecasts has been steadily improving over the past  decades (c.f. ~\citet{bauer2015quiet, alley2019advances}). The earliest  dynamically-modeled numerical weather forecast for a single point was computed using a slide rule and table of logarithms by Lewis Fry Richardson in 1922~\citep{richardson2007weather} and took six weeks to compute a 6-hour forecast of the atmosphere. By the 1950s, early electronic computers greatly improved the speed of forecasting, allowing operational forecasts to be calculated fast enough to be useful for future prediction.  In addition to better computing capabilities, improvements in weather forecasting have been achieved through better parameterization of small-scale processes through deeper understanding of their physics and higher-quality atmospheric observations. The latter has resulted in improved model initializations via data assimilation. 

There is now increasing interest around developing data-driven Deep Learning (DL) models for weather forecasting owing to their orders of magnitude lower computational cost as compared to state-of-the-art NWP models \citep{schultz2021can,balaji2021climbing,irrgang2021towards,reichstein2019deep}. Many studies have attempted to build data-driven models for forecasting the large-scale circulation of the atmosphere, either trained on climate model outputs, general circulation models (GCM) \citep{scher2018predicting,scher2019weather, chattopadhyay2019analog}, reanalysis products \citep{weyn2019can,weyn2020improving,weyn2021sub,rasp2020weatherbench,rasp2021data,rasp2020purely,chattopadhyay2021towards,arcomano2020machine,chantry2021opportunities,gronquist2021deep}, or a blend of climate model outputs and reanalysis products \citep{rasp2021data}.

Data-driven models have great potential to improve weather predictions by overcoming model biases present in NWP models and by enabling the generation of large ensembles at low computational cost for probabilistic forecasting and data assimilation. By training on reanalysis data or observations, data-driven models can avoid limitations that exist in NWP models \citep{schultz2021can,balaji2021climbing}, such as biases in convection parameterization schemes that strongly affect precipitation forecasts. Once trained, data-driven models are orders of magnitude faster than traditional NWP models in generating forecasts via inference, thus enabling the generation of very large ensembles~\citep{chattopadhyay2021towards,weyn2021sub}. 

In this regard, \citet{weyn2021sub} have shown that large data-driven ensembles improve subseasonal-to-seasonal (S2S) forecasts over operational NWP models that can only incorporate a small number of ensemble members. Furthermore, a large ensemble helps improve data-driven predictions of extreme weather events in short- and long-term forecasts ~\citep{chattopadhyay2019analog}. 

Most data-driven weather models, however, use low-resolution data for training, usually at the $5.625^{\circ}$ resolution as in \citet{rasp_2020_resnet} or $2^{\circ}$ as in \citet{weyn2020improving}. These prior attempts have achieved good results on forecasting some of the coarse, low-resolution atmospheric variables. However, the coarsening procedure leads to the loss of crucial, fine-scale physical information. For data-driven models to be truly impactful, it is essential that they generate forecasts at the same or greater resolution than current state-of-the-art numerical weather models, which are run at $\approx 0.1^{\circ}$ resolution. Forecasts at $5.625^{\circ}$ spatial resolution, for instance, result in a mere $32 \times 64$ pixels grid representing the entire globe. Such a forecast is not able to resolve features smaller than $\approx$ 500 km. Such coarse forecasts fail to account for the important effects of small-scale dynamics on the large scales and the impact of topographic features such as mountain ranges and lakes on small-scale dynamics. This limits the practical utility of low-resolution forecasts. While low-resolution forecasts may be justified for variables that do not possess a lot of small-scale structures, such as the geo-potential height at 500 hPa ($Z_{500}$), higher-resolution data (e.g., at $0.25^{\circ}$ resolution) can substantially improve the predictions of data-driven models for variables like low-level winds ($U_{10}$ and $V_{10}$) that have complex fine-scale structures. Moreover, high-resolution models can resolve the formation and dynamics of high-impact extreme events such as tropical cyclones, which would otherwise be inadequately represented on a coarser grid.

{\bf Our approach: } We develop FourCastNet, a Fourier-based neural network forecasting model, to generate global data-driven forecasts of key atmospheric variables at a resolution of $0.25\degree$, which corresponds to a spatial resolution of roughly 30 km $\times$ 30 km near the equator and a global grid size of $720 \times 1440$ pixels. This allows us, for the first time, to make a direct comparison with the high-resolution Integrated Forecasting System (IFS) model of the European Center for Medium-Range Weather Forecasting (ECMWF).

\begin{figure}
    \centering
    \includegraphics[width = 1.0\textwidth]{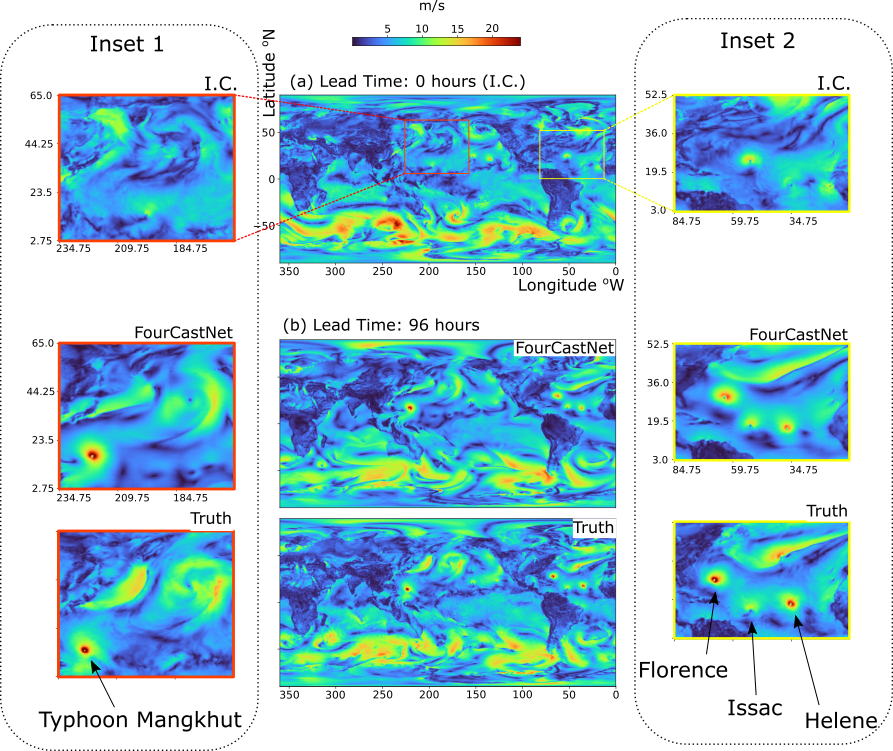}
    \caption{Illustrative example of a global near-surface wind forecast generated by FourCastNet over the entire globe at a resolution of $0.25^{\degree}$. To prepare this figure, we initialized FourCastNet with an initial condition from the out-of-sample test dataset with the calendar timestamp September 8, 2018 at 00:00 UTC. Starting from this initial condition, the model was allowed to run freely for 16 time-steps of six hours each in inference mode (Figure~\ref{fig:afno}(d)) corresponding to a 96-hour forecast. Panel (a) shows the wind speed at model initialization. Panel (b) shows the model forecasts at forecast lead time of 96 hours (upper panel) and the corresponding true wind speeds at that time (lower panel). FourCastNet is able to forecast the wind speeds 96 hours in advance with remarkable fidelity and correct fine-scale features. The forecast accurately captures the formation and track of Super Typhoon Mangkhut that begins to form at roughly $10^{\degree} N$, $210^{\degree} W$ (see Inset 1). Further, the model captures the intensification and track of the typhoon over a period of four days. During the period of this forecast, the model reveals three named hurricanes (Florence, Issac, and Helene) forming in the Atlantic Ocean and approaching the eastern coast of North America (see Inset 2). Further discussion of hurricane forecasts using FourCastNet is provided in Section~\ref{sec:hurricane} and Appendix \ref{app:MSLP_hurricane}.}
    \label{fig:mangkhut}
\end{figure}

Figure~\ref{fig:mangkhut} shows an illustrative global near-surface wind speed forecast at a 96-hour lead time generated using FourCastNet. We highlight key high-resolution details that are resolved and accurately tracked by our forecast, including Super Typhoon Mangkhut and three named cyclones heading towards the eastern coast of the United States (Florence, Issac, and Helene).

FourCastNet is about 45,000 times faster than traditional NWP models on a node-hour basis. This orders of magnitude speedup, along with the unprecedented accuracy of FourCastNet at high resolution, enables inexpensive generation of extremely large ensemble forecasts. This dramatically improves probabilistic weather forecasting. Massive large-ensemble forecasts of events such as hurricanes, atmospheric rivers, and extreme precipitation can be generated in seconds using FourCastNet. This could lead to better-informed disaster response. Furthermore, FourCastNet's reliable, rapid, and cheap forecasts of near-surface wind speeds can improve wind energy resource planning at onshore and offshore wind farms. The energy required to train FourCastNet is approximately equal to the energy required to generate a 10-day forecast with 50 ensemble members using the IFS model. Once trained, however, FourCastNet uses about 12,000 times less energy to generate a forecast than the IFS model. We expect FourCastNet to be only trained once; the energy consumption of subsequent fine tuning is negligible.

FourCastNet uses a Fourier transform-based token-mixing scheme \citep{guibas2021adaptive} with a vision transformer (ViT) backbone~\citep{dosovitskiy2021image}. This approach is based on the recent Fourier neural operator that learns in a resolution-invariant manner and has shown success in modeling challenging partial differential equations (PDE) such as fluid dynamics~\citep{li2021fourier}. We chose a ViT backbone since it is capable of modeling long-range dependencies well. Combining ViT with Fourier-based token mixing yields a state-of-the-art high-resolution model that resolves fine-grained features and scales well with resolution and size of dataset. This approach enables training high-fidelity data-driven models at truly unprecedented resolution.\footnote{We estimate that FourCastNet could be trained on currently available GPU hardware in about two months with 40 years of global 5-km data, if such data were available.}

In summary, FourCastNet makes four unprecedented contributions to data-driven weather forecasting: 
\begin{enumerate}[label=\arabic{enumi}.,ref= \arabic{enumi}, leftmargin=*]

\item FourCastNet predicts, with unparalleled accuracy at forecast lead times of up to one week, challenging variables such as surface winds and precipitation. No deep learning (DL) model thus far has attempted to forecast surface winds on global scales. Additionally, DL models for precipitation on global scales have been inadequate for resolving small-scale structures. This has important implications for disaster mitigation and wind energy resource planning. 

\item FourCastNet has eight times greater resolution than state-of-the-art DL-based global weather models. Due to its high resolution and accuracy, FourCastNet resolves extreme events such as tropical cyclones and atmospheric rivers that have been inadequately represented by prior DL models owing to their coarser grids.

\item FourCastNet's predictions are comparable to the IFS model on metrics of Root Mean Squared Error (RMSE) and Anomaly Correlation Coefficient (ACC) at lead times of up to three days. After, predictions of all modeled variables lag close behind IFS at lead times of up to a week. Whereas the IFS model has been developed over decades, contains greater than 150 variables at more than 50 vertical levels in the atmosphere, and is guided by physics, FourCastNet models 20 variables at five vertical levels, and is purely data driven. This comparison points to the enormous potential of data-driven modeling in complementing and eventually replacing NWP.

\item FourCastNet's reliable, rapid, and computationally inexpensive forecasts facilitate the generation of very large ensembles, thus enabling estimation of well-calibrated and constrained uncertainties in extremes with higher confidence than current NWP ensembles that have at most approximately 50 members owing to their high computational cost. Fast generation of 1,000-member ensembles dramatically changes what is possible in probabilistic weather forecasting, including improving reliability of early warnings of extreme weather events and enabling rapid assessment of their impacts.
\end{enumerate}

\section{Training Methods}\label{sec:training}
 The ECMWF provides a publicly available, comprehensive dataset called ERA5~\citep{hersbach2020era5} which consists of hourly estimates of several atmospheric variables at a latitude and longitude resolution of $0.25\degree$ from the surface of the earth to roughly 100 km altitude from 1979 to the present day. ERA5 is an atmospheric reanalysis~\citep{kalnay1996ncep} dataset and is the result of an optimal combination of observations from various measurement sources and the output of a numerical model using a Bayesian estimation process called data-assimilation~\citep{kalnay2003atmospheric}. The dataset is essentially a reconstruction of the optimal estimate of the observed history of the Earth's atmosphere. We use the ERA5 dataset to train FourCastNet. While the ERA5 dataset has several prognostic variables available at 37 vertical levels with an hourly resolution, computational and data limitations along with other operational considerations for DL models restricts our choice, based on physical reasoning, to a subset of these available variables to train our model on. 

In this work, we focus on forecasting two important and challenging atmospheric variables, namely, (1) the wind velocities at a distance of 10m from the surface of the earth and (2) the 6-hourly total precipitation. There are a few reasons for our focus on these variables. First, surface wind velocities and precipitation require high-resolution models to resolve and forecast accurately because they contain and are influenced by many small-scale features. Due to computational and model architectural limitations, previous efforts in DL-based weather prediction have not been able to produce global forecasts for these variables at full ERA5 resolution. Near-surface wind velocity forecasts have a tremendous amount of utility due to their key role in planning energy storage, grid transmission, and other operational considerations at on-shore and off-shore wind farms. As we show in Section~\ref{sec:hurricane}, near-surface wind forecasts (along with wind forecasts above the atmospheric boundary layer) can help track extreme wind events such as hurricanes and can be used for disaster preparedness. Our second focus is on forecasting total precipitation where DL models can potentially show great promise. NWP models, such as the operational IFS, have several parameterization schemes to tractably forecast precipitation and since neural networks are known to have impressive capabilities at deducing parameterizations from high-resolution observational data, they are well-suited for this task. 

Although we focus on forecasting near-surface wind-speed and precipitation, our model also forecasts with remarkable accuracy several other variables. In our forecast, we include the geopotential height, temperature, wind velocity, and relative humidity at a few different vertical levels, a few near-surface variables such as surface pressure and mean sea-level pressure as well as the integrated total column of water vapor.

\subsection{FourCastNet: Model Description}

 To produce our high-resolution forecasts, we choose the Adaptive Fourier Neural Operator (AFNO) model ~\citep{guibas2021adaptive}. This particular neural network architecture is appealing as it is specifically designed for {\it high-resolution} inputs and synthesizes several key recent advances in DL into one model. Namely, it combines the Fourier Neural Operator (FNO) learning approach of \cite{li2021fourier}, which has been shown to  perform well in modeling challenging PDE systems, with a powerful ViT backbone.
 
The vision transformer (ViT) architecture and its variants have emerged as the state-of-the-art in computer vision over the previous years, showing remarkable performance on a number of tasks and scaling well with increased model and dataset sizes. Such performance is attributed mainly to the multi-head self-attention mechanism in these networks, which allows the network to model interactions between features (called tokens in ViT representation terms) globally at each layer in the network. However, spatial mixing via self-attention is quadratic in the number of tokens, and thus quickly becomes infeasible for high-resolution inputs.

Several ViT variants with reduced computational complexity have been proposed, with various alternate mechanisms for spatial token mixing employed in each. However, the AFNO model is unique in that it frames the mixing operation as continuous global convolution, implemented efficiently in the Fourier domain with FFTs, which allows modeling dependencies across spatial and channel dimensions flexibly and scalably. With such a design, the spatial mixing complexity is reduced to $\mathcal{O}(N\log N)$, where $N$ is the number of image patches or tokens. This scaling allows the AFNO model to be well-suited to high-resolution data at the current $0.25^{\degree}$ resolution considered in this paper as well as potential future work at an even higher resolution. In the original FNO formulation, the operator learning approach showed impressive results solving turbulent Navier-Stokes systems, so incorporating this into a data-driven atmospheric model is a natural choice.

Given the general popularity of convolutional network architectures, and particularly their usage in previous works forecasting ERA5 variables \citep{rasp_2020_resnet, weyn2020improving}, it is worth contrasting our AFNO model with these more conventional architectures. For one, the ability of AFNO to scale well with resolution yields immediate practical benefits -- at our 720x1440 resolution, the FourCastNet model memory footprint is about 10GB with a batch size of 1. To contrast this, we can look at the 19-layer ResNet architecture from a prior result on WeatherBench \citep{rasp_2020_resnet}, which was trained at a very coarse resolution (32$\times$64 pixels). Naively transferring this architecture to our dataset and training at 720$\times$1440 resolution would require 83GB for a batch size of 1. This is prohibitive, and is compounded by the fact that it is somewhat of a lower bound -- with order-of-magnitude increases in resolution, a convolution-based network's receptive field would similarly need to grow via the addition of even more layers.

Beyond practical considerations, our preliminary non-exhaustive experiments suggested that convolutional architectures showed poor performance on capturing small scales over many time steps in auto-regressive inference. These observations along with our knowledge of the current state of the art for high-resolution image processing in image de-noising, super-resolution and de-blurring are a strong motivation for our choice of a ViT architecture over a convolutional architecture.

While we refer the reader to the original AFNO paper \citep{guibas2021adaptive} for more details, we briefly describe the flow of computation in our model here. First, the input variables on the $720\times 1440$ lat-lon grid are projected to a 2D grid ($h \times w$) of patches (with a small patch size $p\times p$, where e.g., $p=8$), with each patch represented as a $d$-dimensional token. Then, the sequence of patches are fed, along with a positional encoding, to a series of AFNO layers. Each layer, given an input tensor of patches $X \in \mathbb{R}^{h\times w \times d}$, performs spatial mixing followed by channel mixing. Spatial mixing happens in the Fourier domain as follows:

\begin{itemize}

    \item[\textbf{Step 1}.] Transform tokens to the Fourier domain with
    \begin{equation}
        z_{m,n} = [\mathrm{DFT}(X)]_{m,n},
    \end{equation}
    where $m,n$ index the patch location and DFT denotes a 2D discrete Fourier transform.
    
    \item[\textbf{Step 2}.] Apply token weighting in the Fourier domain, and promote sparsity with a Soft-Thresholding and Shrinkage operation as
    \begin{equation}
        \tilde{z}_{m,n} = S_{\lambda} ( \mathrm{MLP}(z_{m,n})),
    \end{equation}
    where $S_{\lambda}(x) = \mathrm{sign}(x) \max(|x| - \lambda, 0)$ with the sparsity controlling parameter $\lambda$, and MLP() is a 2-layer multi-layer perceptron with block-diagonal weight matrices which are shared across all patches.
    
    \item[\textbf{Step 3}.] Inverse Fourier to transform back to the patch domain and add a residual connection as
    \begin{equation}
        y_{m,n} = [\mathrm{IDFT}(\tilde{Z})]_{m,n} + X_{m,n}.
    \end{equation}
    
\end{itemize}

\begin{figure}
    \centering
    \includegraphics[width = \textwidth]{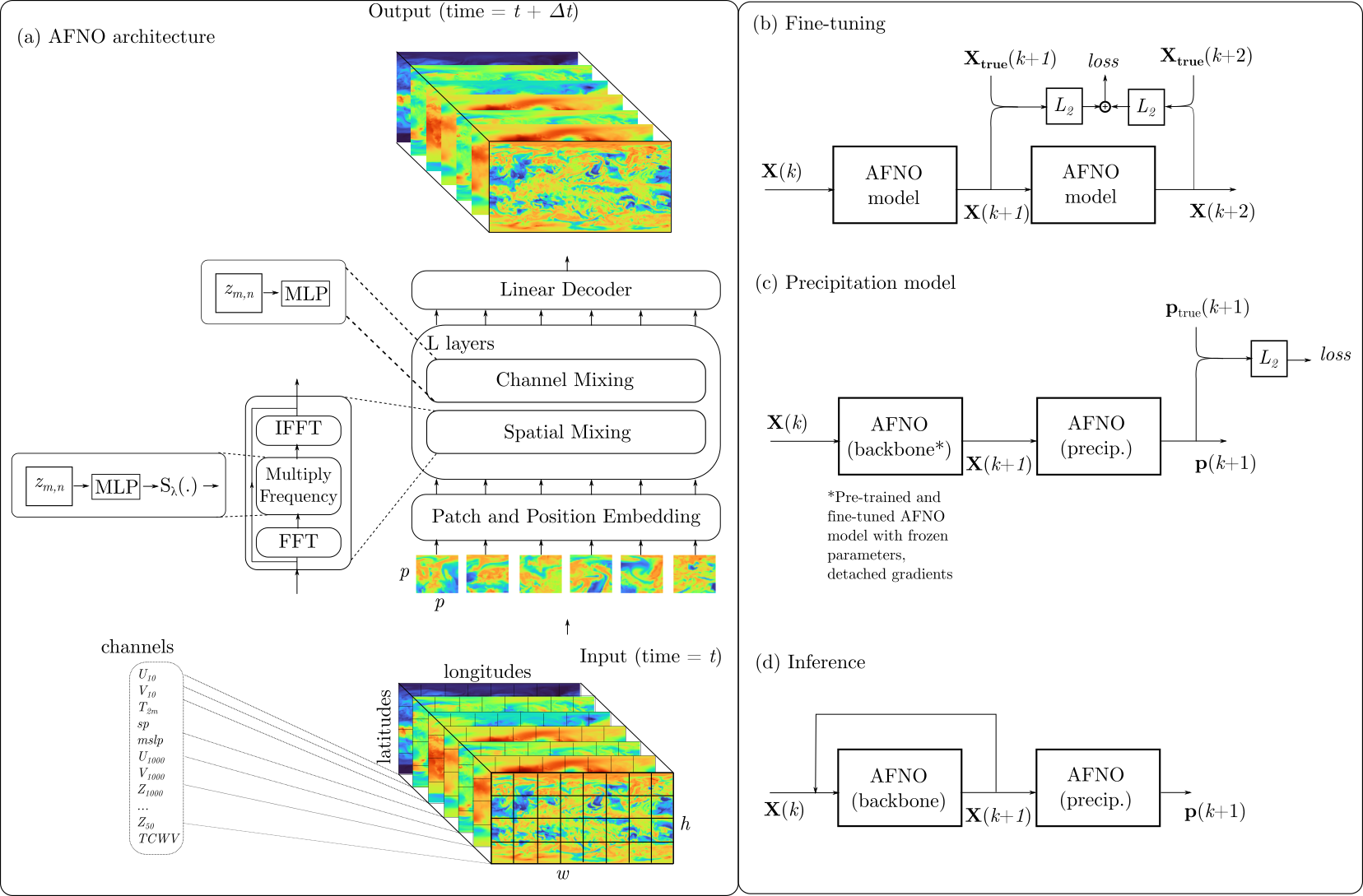}
    \caption{(a) The multi-layer transformer architecture that utilizes the Adaptive Fourier Neural Operator with shared MLP and frequency soft-thresholding for spatial token mixing. The input frame is first divided into a $h \times w$ grid of patches, where each patch has a small size $p \times p \times c$. Each patch is then embedded in a higher dimensional space with high number of latent channels and position embedding is added to form a sequence of tokens. Tokens are then mixed spatially using AFNO, and subsequently for each token the latent channels are mixed. This process is repeated for $L$ layers, and finally a linear decoder reconstructs the patches for the next frame from the final embedding. The right-hand panels describe the FourCastNet model's additional training and inference modes: (b) two-step fine-tuning, (c) backbone model that forecasts the 20 variables in Table~\ref{tab:table_era5_variables} with secondary precipitation diagnostic model (note that $\textbf{p}(k+1)$ denotes the 6 hour accumulated total precipitation that falls between $k+1$ and $k+2$ time steps) (d) forecast model in free-running autoregressive inference mode.}
    \label{fig:afno}
\end{figure}

\subsection{Training}

While our primary interest lies in forecasting the surface wind velocities and precipitation, the complex atmospheric system contains strong nonlinear interactions across several variables such as temperatures, surface pressures, humidity, moisture content from the surface of the earth to the stratosphere, etc. In order to model these interactions, we choose a few variables (Table~\ref{tab:table_era5_variables}) to represent the instantaneous state of the atmosphere. These variables are specifically chosen to model important processes that influence low-level winds and precipitation. As such, we treat all the prognostic variables equally and the model architecture or optimization scheme does not afford special treatment to any of the prognostic variables.

Each of the variables in Table~\ref{tab:table_era5_variables} is re-gridded from a Gaussian grid to a regular Euclidean grid using the standard interpolation scheme provided by the Copernicus Climate Data Store (CDS) Application Programming Interface (API). Following the re-gridding process, each of the 20 variables is represented as a 2D field of shape ($721 \times 1440$) pixels. Thus, a single training data point at an instant in time containing all 20 variables is represented by a tensor of shape ($721 \times 1440 \times 20$). While the ERA5 dataset is available at a temporal resolution of 1 hour, we choose to sub-sample the dataset and use snapshots spaced 6 hours apart to train our model. Within each 24 hour day, we choose to sample the 20 variable subset of the ERA5 dataset at $0000$ hrs, $0600$ hrs, $1200$ hrs and $1800$ hrs. We divide the dataset into three sets, namely training, validation and out-of-sample testing datasets. The training dataset consists of data from the year 1979 to 2015 (both included). The validation dataset contains data from the years 2016 and 2017. The out-of-sample testing dataset consists of the years 2018 and beyond.

We collectively denote the modeled variables by the tensor $\vv{X}(k\Delta t)$, where $k$ denotes the time index and $\Delta t$ is the temporal spacing between consecutive snapshots in the training dataset. We will consider the ERA5 dataset as the truth and denote the \textit{true} variables by $\vv{X}_{\text{true}}(k\Delta t)$. With the understanding that $\Delta t$ is fixed at $6$ hours throughout this work, we omit $\Delta t$ in our notation for convenience where appropriate. The training procedure consists of two steps, pre-training and fine-tuning. In the pre-training step, we train the AFNO model using the training dataset in a supervised fashion to learn the mapping from $\vv{X}(k)$ to $\vv{X}(k+1)$. In the fine-tuning step, we start from the previously pre-trained model and optimize the model to predict two time steps, i.e., The model first generates the output $\vv{X}(k+1)$ from the input $\vv{X}(k)$. The model then uses its own output $\vv{X}(k+1)$ as an input and generates the output $\vv{X}(k+2)$. We then compute a training loss by comparing each of $\vv{X}(k+1)$ and $\vv{X}(k+2)$ to the respective ground truth from the training data and use the sum of the two training losses for optimizing the model. In both, the pre-training and fine-tuning steps, the training dataset is used to optimize the model and the validation dataset is used to estimate the model skill during hyper-parameter optimization. The out-of-sample testing dataset is  untouched. The training dataset consists of 54020 samples while the validation dataset contains 2920 samples. We refer to the trained and fine-tuned model as the `\textit{backbone}'. The model is pre-trained using a cosine learning-rate schedule with a starting learning rate $\ell_1$ for 80 epochs. Following the pre-training, the model is fine-tuned for a further 50 epochs using a cosine learning-rate schedule and a lower learning rate $\ell_2$. The precipitation model (described in Section~\ref{sec:precip_model}) is then added to the trained backbone and trained for 25 epochs using a cosine learning rate schedule  with an initial learning rate $\ell_3$. The learning rates and other training hyperparameters are provided in Table~\ref{tab:hyperparameters} in Appendix \ref{app:hyperparams}. The end to end training takes about 16 hours wall-clock time on a cluster of 64 Nvidia A100 GPUs.

\begin{table}
	\centering
	\begin{tabular}{ll}
	\toprule
	Vertical Level & Variables \\
	\hline
	Surface & $U_{10}$, $V_{10}$, $T_{2m}$, $sp$, $mslp$ \\
	$1000hPa$ & $U$, $V$, $Z$ \\
	$850hPa$ & $T$, $U$, $V$, $Z$, $RH$ \\
	$500hPa$ & $T$, $U$, $V$, $Z$, $RH$ \\
	$50hPa$ & $Z$ \\
	Integrated & $TCWV$\\
	\bottomrule
	\end{tabular}
	\caption{Prognostic Variables modeled by the DL model. Abbreviations are as follows. $U_{10}$ ($V_{10}$): zonal (meridonal) wind velocity 10m from the surface; $T_{2m}$: Temperature at 2m from the surface; $T$, $T$, $V$, $Z$. $RH$: Temperature, zonal velocity, meridonal velocity, geopotential, relative humidity respectively at specified vertical level; $TCWV$: Total Column Water Vapor.}
	\label{tab:table_era5_variables}
\end{table}

\subsection{Precipitation Model}
\label{sec:precip_model}
The total precipitation ($TP$) in the ERA5 re-analysis dataset is a variable that represents the the accumulated liquid and frozen water that falls to the Earth's surface through rainfall and snow. It is defined in units of length as the depth of water that would accumulate if spread evenly over a unit grid box of the model. Compared to the variables handled by our backbone model, $TP$ exhibits certain features that complicate the task of forecasting it---the probability distribution of $TP$ is strongly peaked at zero with a long tail towards positive values. Hence, $TP$ exhibits more sparse spatial features than the other prognostic variables. In addition, $TP$ does not have significant impact on the variables that guide the dynamical evolution of the atmosphere (e.g. winds, pressures, and temperatures), and capturing it accurately in NWP involves complex parameterizations for processes like phase changes.

For these reasons, we treat the total precipitation ($TP$) as a diagnostic variable and denote it by $\vv{p}(k\Delta t)$. Total precipitation is not included in the 20 variable dataset used to train the backbone model\footnote{This approach is similar to previous work \citep{rasp_2020_resnet}, which trained a separate model for precipitation than for the other atmospheric variables.}. Rather, we train a separate AFNO model to diagnose $TP$ using the outputs of the backbone model, as indicated in Figure~\ref{fig:afno}(c). This approach decouples the difficulties of modeling precipitation (which typically deteriorates in accuracy fairly quickly) from the general task of forecasting the atmospheric state. In addition, once trained, our diagnostic $TP$ model could potentially be used in conjunction with other forecast models (either traditional NWP or data-driven forecasts).

The model used to diagnose precipitation from the output of the backbone has the same base AFNO architecture, with an additional 2D convolutional layer (with periodic padding) and a ReLU activation as the last layer, used to enforce non-negative precipitation outputs. Since the backbone model makes predictions in 6-hour increments, we train our diagnostic precipitation model to predict the 6-hourly accumulated total precipitation (rather than the 1 hour precipitation in the raw ERA5 data). This also enables easy comparison with the IFS model, which is archived in 6-hour increments and thus also predicts 6-hourly accumulated precipitation. Following \citep{rasp2020weatherbench}, we additionally log-transform the precipitation field: $\tilde{TP} = \log{(1 + TP/\epsilon)}$, with $\epsilon = \num{1E-5}$. Since total precipitation values are highly sparse, this transformation discourages the network from predicting zeros and ensures a less skewed distribution of values. For any comparisons with the IFS model or ERA5 ground truth, we transform $TP$ back to units of length.

\subsection{Inference}
We generate forecasts of the core atmospheric variables in Table~\ref{tab:table_era5_variables} and the total precipitation by using our trained models in autoregressive inference mode as shown in Figure~\ref{fig:afno}(d). The model is initialized with an initial condition ($\vv{X}_{\text{true}}(j)$) from the year 2018~\footnote{The year 2018 was chosen from the out-of-sample dataset due to ready availability of IFS forecasts for that year from the TIGGE archive.} out-of-sample held out dataset for $N_f$ different initial conditions and allowed to freely run iteratively for $\tau$ time-steps to generate forecasts $\lbrace \vv{X}_{\text{pred}}(j + i\Delta t) \rbrace_{i=1}^\tau$. The initial conditions $\vv{X}_{\text{true}}(j)$ are spaced apart by $D$ days based on a rough estimate of the temporal de-correlation time for each of the variables being forecast. The value of $D$ and $N_f$ is thus different for each of the forecast variables and listed in Table~\ref{tab:IC} of Appendix \ref{app:hyperparams} unless otherwise specified. We also use the IFS forecasts for the year 2018 from The International Grand Global Ensemble (TIGGE) archive for comparative analysis. The archived IFS forecasts, with initial conditions matching the times of corresponding initial conditions for the FourCastNet model forecast, are used for comparing our model's accuracy to that of the IFS model.

\section{Results}
\label{sec:results}

Figure~\ref{fig:mangkhut} qualitatively shows the forecast skill of our FourCastNet model on forecasting the surface wind speeds over the entire globe at a resolution of $0.25^{\degree}$-lat-long. The wind speeds are computed as the magnitude of the surface wind velocity using the zonal and meridonal components of the wind velocity i.e., $\sqrt{\left( U_{10}^2 + V_{10}^2 \right)}$ To prepare this figure, we initialized the FourCastNet model with an initial condition from the out-of-sample test dataset. Starting from this initial condition, the model was allowed to run freely for 16 time-steps in inference mode (Figure~\ref{fig:afno}(d)). The calendar time-stamp of the initial condition used to generate this forecast was September 8, 2018 at 00:00 UTC. Figure~\ref{fig:mangkhut}(a) shows the wind speed at model initialization. Figure~\ref{fig:mangkhut}(b) shows the model forecasts at a lead time of 96 hours (upper-panel) and the corresponding true wind speeds at that time (lower-panel). We note that the FourCastNet model is able to forecast the wind speeds upto 96 hours in advance with remarkable fidelity with correct fine-scale features. Notably, this figure illustrates the forecast of the formation and track of a super-typhoon named Mangkhut that is beginning to form in the initialization frame at roughly $10^{\degree} N$ latitude, $210^{\degree} W$ longitude. The model qualitatively tracks with remarkable fidelity the intensification of the typhoon and its track over a period of 4 days. Also of note are three simultaneous named hurricanes (Florence, Issac and Helene) forming in the Atlantic ocean and approaching the eastern coast of North America during the period of this forecast. The FourCastNet model appears to be able to forecast the formation and track of these phenomena remarkably well. We provide a further discussion of hurricane forecasts with a few quantitative results and case studies in Section~\ref{sec:hurricane} and Appendix \ref{app:MSLP_hurricane}.

\begin{figure}
    \centering
    \includegraphics[width  = 0.9\textwidth]{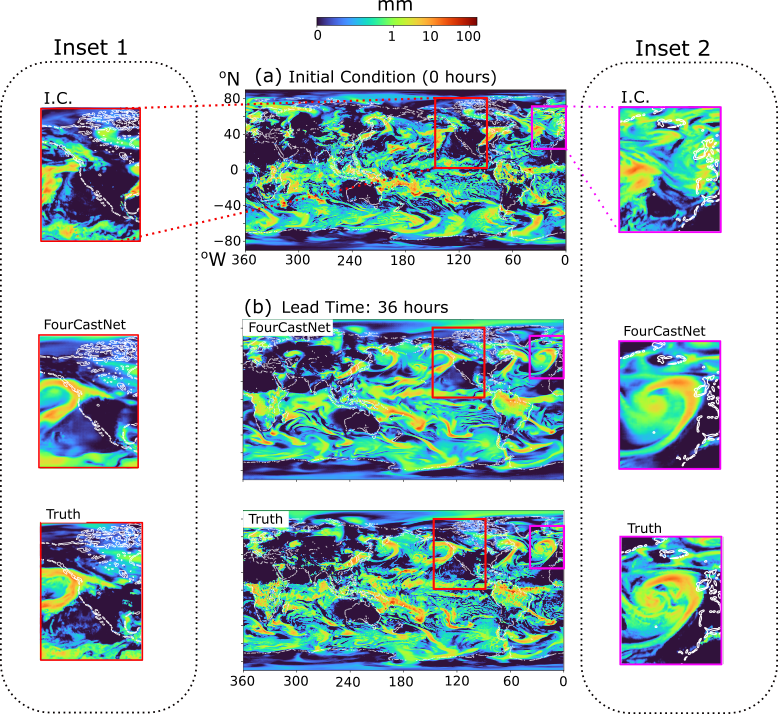}
    \caption{Illustration of a global Total Precipitation (TP) forecast using the FourCastNet model. Land-sea borders are shown using a thin white trace. For ease of visualization, the precipitation field is plotted as a log-transformed field in all panels. Panel (a) shows the TP fields at the time of forecast initialization. Panel (b) shows the TP forecast generated by the FourCastNet model (upper panel) over the entire globe at $0.25^{\degree}$-lat-long resolution with the corresponding truth (lower panel). Inset 1 shows the I.C., forecast and true precipitation fields at a lead time of 36 hours over a local region along the western coast of the United States. This highlights the ability of the FourCastNet model to resolve and predict localized regions of high precipitation, in this case due to an atmospheric river. Inset 2 shows the I.C., forecast, and true precipitation fields near the coast of the U.K. and highlights an extreme precipitation event due to an extra-tropical cyclone that is predicted very well by the FourCastNet model. The precipitation is diagnosed from the FourCastNet predicted prognostic variables as described in Figure~\ref{fig:afno}(d). The calendar time-stamp of the initial condition used to generate this forecast was 00:00 UTC on April 4, 2018. The high-resolution FourCastNet model demonstrates excellent skill in capturing small scale features that are key to precipitation forecasting.}
    \label{fig:precip}
\end{figure}

In Fig~\ref{fig:precip}, we show the forecast skill of our model in diagnosing total precipitation over the entire globe. Using the free running FourCastNet model predictions (from above) for the 20 prognostic variables as input to the precipitation model, we diagnose total precipitation at the same time steps. Fig~\ref{fig:precip}(a) shows the precipitation at the initial time, Fig~\ref{fig:precip}(b) shows the model predictions at lead time 36 hours along with the corresponding ground truth. The inset panels show the precipitation fields over a local region along the western coast of the United States, highlighting the ability of the FourCastNet model to resolve and forecast localized areas of high precipitation with remarkable accuracy. Forecasting precipitation is known to be an extremely difficult task due to its intermittent and stochastic nature. Despite these challenges, we observe that the FourCastNet diagnosis shows excellent skill in capturing short-term high-resolution precipitation features, which can have significant impact in predicting extreme events. We also note that this is the first time a DL model has been successfully utilized to provide competitive precipitation diagnosis at this scale.

\subsection{Hurricanes}\label{sec:hurricane}
In this section, we explore the potential utility of developing DL models for forecasting hurricanes, a category of extreme events with tremendous destructive potential. A rapidly available, computationally inexpensive atmospheric model that could could forewarn the possibility of hurricane formation and track the path of the hurricane would be of great utility for mitigating loss of life and property damage. As the stakes for mis-forecasting such extreme weather phenomena are very high, more rigorous studies need to be undertaken before DL can be considered a mature technology to forecast hurricanes. The results herein should be considered a preliminary and exploratory dive for inspiring future research into the potential of DL models to provide valuable models of this phenomenon. Prior to this work, DL models were trained on data that was too coarse and thus incapable of resolving atmospheric variables finely enough (see Appendix \ref{app:MSLP_hurricane} for an illustration). Prior models could not generate accurate predictions of wind speed and other important prognostic variables with long enough forecast lead times to consider hurricane forecasts. Our model has reasonably good resolution and generates accurate medium-range forecasts of variables that allow us to track the generation and path of hurricanes. For a case-study we consider a hurricane that occurred in 2018 (a year that is part of our out-of-sample dataset), namely hurricane Michael.

Michael was a category 5 hurricane on the Saffir -Simpson Hurricane Wind Scale that made landfall in Florida causing catastrophic damage~\citep{hcmichael}. Michael started as a tropical depression around October 7, 2018. Within a day, the depression intensified into a hurricane. After undergoing rapid intensification in the gulf of Mexico, Michael reached category 5 status. Soon after, Michael made landfall in Florida on October 10, 2018. Thus within a short period of roughly 72 hours, Michael went from a tropical depression to a category 5 hurricane to landfall.

\begin{figure}
    \centering
    \includegraphics[width = \textwidth]{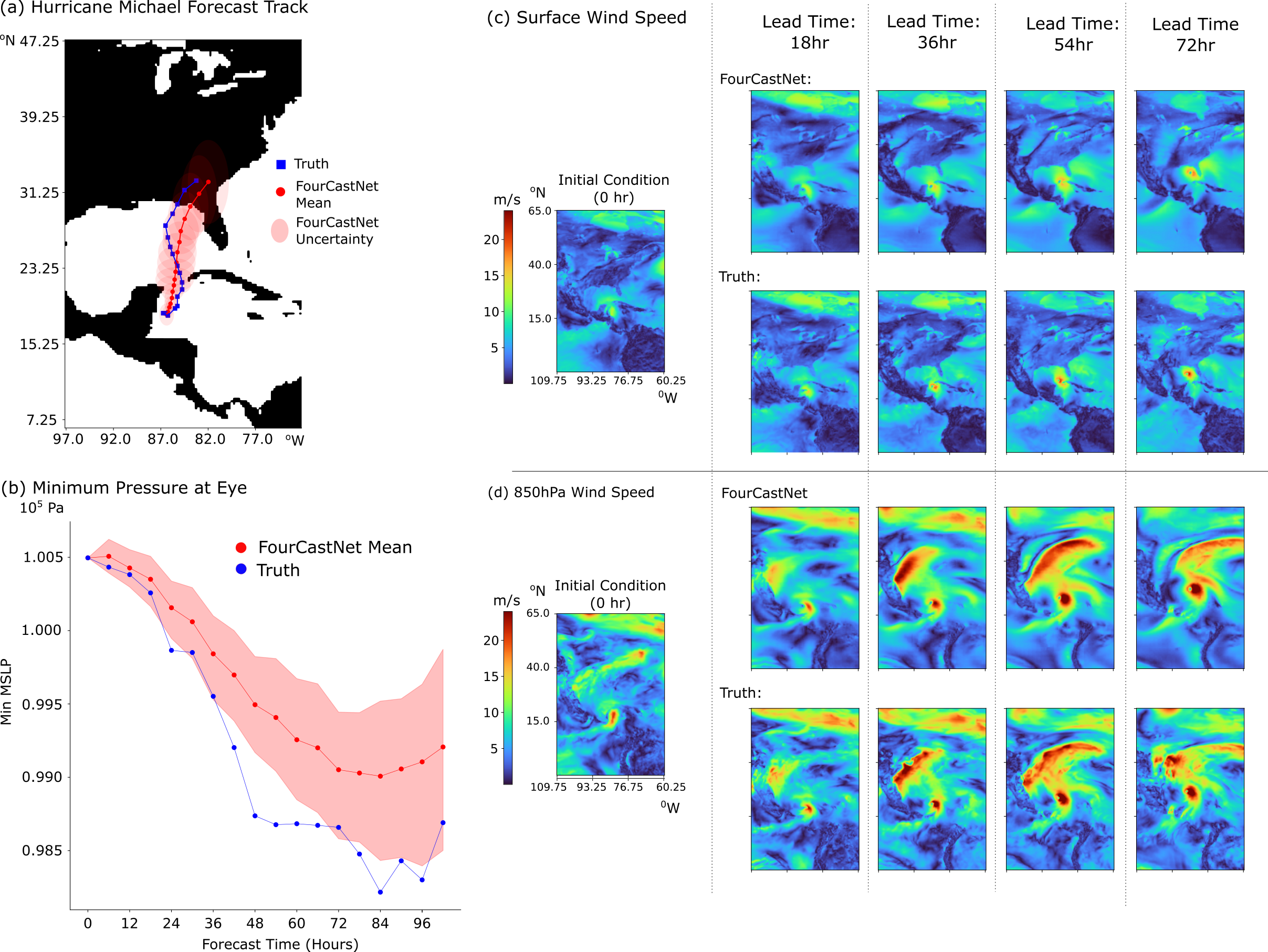}
    \caption{The FourCastNet model has excellent skill on forecasting fine-scale, rapidly changing variables relevant to a hurricane forecast. As an illustrative example, we have chosen Hurricane Michael which underwent rapid intensification during the course of its four day trajectory. Panel (a) shows the mean position of the minima of Mean Sea Level Pressure (indicating the eye of hurricane Michael) as forecast by a 100 member ensemble forecast using FourCastNet (red circles) and the corresponding ground truth according to ERA5 reanalysis (blue squares) for 108 hours starting from the initial condition at 00:00 hours on October 7, 2018 UTC. To generate an ensemble forecast, the initial condition was perturbed with Gaussian noise as described in Section~\ref{sec:ensemble} and 100 forecast trajectories were computed. The shaded ellipses have a width and height equal to the 90th percentile spread of the longitudinal and latitudinal positions respectively of the hurricane eye as indicated by the MSLP minima in the 100-member FourCastNet ensemble. Panel (b) shows the minimum MSLP at the eye of hurricane Michael as forecast by FourCastNet (red filled circles) along with the corresponding true minimum from the ERA5 reanalysis (blue filled circles). The red shaded region shows the 90 percent confidence region in the 100-member ensemble forecast. Panels (c) and (d) respectively show the surface wind speed and 850hPa wind speed predictions at lead times of 18 hours, 36 hours, 54 hours and 72 hours generated by FourCastNet along with the corresponding true wind speeds at those times. The surface wind speed and the 850hPa speed in the initial condition (Oct. 7, 2018 00:00 UTC) that was used to initialize this forecast is shown in the leftmost column. Collectively, the minimum MSLP tracks, surface wind speed and the 850hPa wind speed forecasts show the formation, intensification and path of Hurricane Michael as it goes from a tropical depression to a category 5 hurricane with landfall on the west coast of Florida.}
    \label{fig:hcm}
\end{figure}

We use our trained model as described in Section~\ref{sec:training} (with no further changes) to study the potential of our model for forecasting the formation, rapid intensification and tracking of hurricane Michael. The FourCastNet model is capable of rapidly generating large ensemble forecasts. We start from the initial condition at the calendar time 00:00 hours on October 7, 2018 UTC. The initial condition was perturbed with Gaussian noise to generate an ensemble of $E=100$ perturbed initial conditions. We provide further discussion of ensemble forecasting using FourCastNet in Section~\ref{sec:ensemble}. Figure~\ref{fig:hcm} shows the track of the hurricane and the intensification as forecast by the 100-member FourCastNet ensemble using the Mean Sea Level Pressure to estimate the eye of the hurricane and the minimum pressure at the eye. Figure~\ref{fig:hcm}(a) shows the mean position of the minima of Mean Sea Level Pressure using a 100 member ensemble forecast generated by FourCastNet (red circles). The corresponding ground truth according to ERA5 reanalysis is indicated on the same plot (blue squares) over a trajectory spanning 108 hours. The shaded ellipses in the figure have a width and height equal to the 90th percentile spread in the longitudinal and latitudinal positions respectively of the hurricane eye as indicated by the MSLP minima in the 100-member FourCastNet ensemble. Figure~\ref{fig:hcm}(b) quantitatively demonstrates that the FourCastNet model is able to predict the intensification of the hurricane as the hurricane eye pressure drops rapidly in the first 72 hours. The minimum MSLP at the eye of hurricane Michael as forecast by FourCastNet is indicated by red circles and the corresponding true minimum from the ERA5 reanalysis is shown by blue circles. The red shaded region shows the region between the first and third quartiles of minimum MSLP in the 100-member ensemble.  While this is an impressive result for a model trained on $0.25^{\degree}$ resolution data, the model fails to fully forecast the extent of the sharp drop in pressure between 36 and 48 hours. We hypothesize that this is likely due to the fact that the current version of the FourCastNet model does not account for a number of convective and radiative processes that would be crucial to such a forecast. Additionally we expect an AFNO model trained on even higher resolution data to improve such a forecast.

 Figures~\ref{fig:hcm}(c),(d) and Fig~\ref{fig:hcm_mslp} in Appendix \ref{app:MSLP_hurricane} respectively provide a qualitative visualization of three prognostic variables that are useful for tracking the formation, intensification and path of a hurricane, namely the wind speed at the surface and at 850hPa level (calculated as the magnitude of the velocity from the meridional and zonal components of the respective velocity -- $U_{10}$, $V_{10}$, $U_{850}$, $V_{850}$ ), and the Mean Sea Level Pressure. We believe there is tremendous potential to improve these forecasts by training even higher resolution DL weather models using the AFNO architecture. 

Our forecasts of the wind speeds and the mean sea level pressure qualitatively match the ground truth remarkably well over a period of 72 hours. Figures ~\ref{fig:hcm}(a),(b), (c), (d), along with \ref{fig:hcm_mslp} in Appendix \ref{app:MSLP_hurricane} clearly show that the DL model is able to forecast the formation, intensification and track of the hurricane from a tropical depression to landfall on the coast of Florida.

Further research is warranted to quantitatively study the potential of our DL model to accurately forecast hurricanes and similar extreme phenomena but these results show great promise in the ability of DL models to aid in the forecasting of one of the most destructive phenomena affecting human life.

\subsection{Atmospheric Rivers}\label{sec:atmospheric_river}

Atmospheric rivers are columns of moisture that are transported by atmospheric circulation currents and carry large amounts of water vapor from the tropics to the extra-tropical regions. They are called `rivers' as they often carry an amount of water equivalent to that of the flow rate of major rivers. Large atmospheric rivers can cause extreme precipitation upon landfall, with the potential to cause flooding and extensive damage. More moderately-sized atmospheric rivers are crucial to the water supply of the western United States. Thus, forecasting atmospheric rivers and their landfall locations is crucial for early warning of flooding in low-lying coastal areas as well as for water resource planning.

\begin{figure}[h]
    \centering
    \includegraphics[width = 0.7\textwidth]{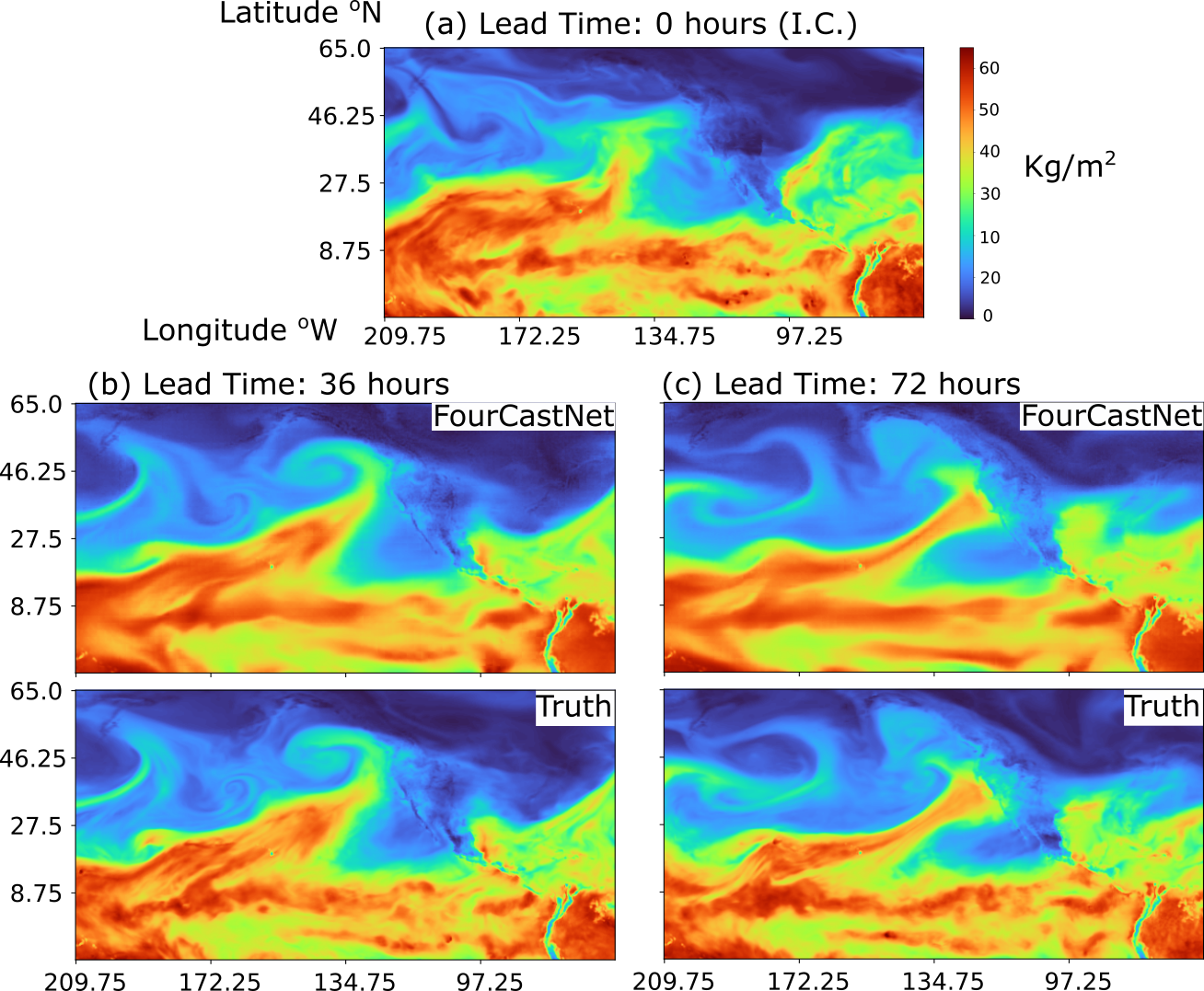}
    \caption{Illustrative example of the utility of the FourCastNet model for forecasting atmospheric rivers. Atmospheric rivers are important phenomena that can cause extreme precipitation and contribute significantly to the supply of precipitable water in several parts of the world. Panels (a)-(c) visualize the Total Column Water Vapor ($TCWV$) in a FourCastNet model forecast initialized at 00:00 UTC on April 4, 2018. Panel (a) Shows the $TCWV$ field in the initial condition that was used to initialize the FourCastNet model. Panels (b) and (c) show the forecasts of the $TCWV$ field produced by the FourCastNet model (top panels) at lead times of 36 and 72 hours respectively along with the corresponding true $TCWV$ fields at those instants of time. The forecast shows an atmospheric river building up and making landfall on the northern California coastline.}
    \label{fig:atmospheric_river}
\end{figure}

Figure~\ref{fig:atmospheric_river} shows the use of our FourCastNet model for predicting the formation and evolution of an atmospheric river (using the Total Column of Water Vapor variable) in April 2018 as it made eventual landfall in Northern California. This type of river which passes through Hawaii is often called the Pineapple Express. Atmospheric rivers show up very clearly in the `Total Column Water Vapor' field that is forecast by the FourCastNet backbone model. The FourCastNet model has very good prediction accuracy for $TCWV$, with ACC$>0.6$ out beyond 8 days as shown by the ACC plot in Figure~\ref{fig:acc_plots_othervars}(d). For this atmospheric river, the FourCastNet model was initialized using an intial condition on April 4, 2018 at 00:00 hours UTC, which we display in Figure~\ref{fig:atmospheric_river}(a). Figures~\ref{fig:atmospheric_river}(b) and \ref{fig:atmospheric_river}(c) show the forecast of the $TCWV$ fields generated by the FourCastNet model (top panels) at a lead time of 36 hours and 72 hours respectively and the corresponding ground truth (bottom panels).

Whie TCWV is a reasonable proxy for atmospheric rivers, we expect future iterations of our model to include Integrated Vapor Transport and Total Column of Liquid Water as additional variables to aid in the forecast of atmospheric rivers.

\subsection{Quantitative Skill of FourCastNet}
\label{sec:quantitative}

We illustrate the forecast skill of our model for $N_f$ initial conditions from the out-of-sample dataset (consisting of the year 2018) and generate a forecast for each initial condition. For each forecast, we evaluate the latitude-weighted Anomaly Correlation Coefficient (ACC) and Root Mean Squared Error (RMSE) for all of the variables included in the forecast. See Appendix \ref{app:acc_rmse_define} for formal definitions of ACC and RMSE. We report the mean ACC and RMSE for each of the variables along with the first and third quartile values of the ACC and RMSE at each forecast time step, to show the dispersion of these metrics over different initial conditions. As a comparison, for the variables listed in Table~\ref{tab:IC}, we also compute the same ACC and RMSE metrics for the corresponding IFS forecast with time-matched initial conditions.

\begin{figure}
	\centering
	\includegraphics[width = 0.9\textwidth]{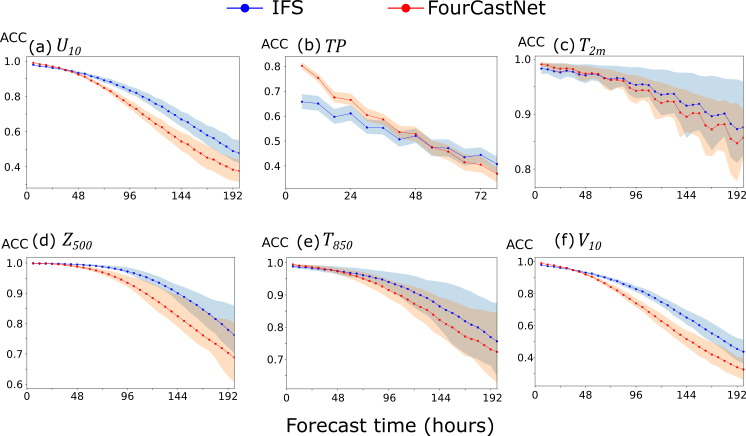}
	\caption{Latitude weighted ACC for the FourCastNet model forecasts (red line with markers) and the corresponding matched IFS forecasts (blue line with markers) averaged over several forecasts initialized using initial conditions in the out-of-sample testing dataset corresponding to the calendar year 2018 for the variables (a) $U_{10}$, (b)  $TP$, (c) $T_{2m}$, (d) $Z_{500}$, (e) $T_{850}$, and (f) $V_{10}$. The ACC values are averaged over $N_f$ initial conditions over a full year with an interval of $D$ days between consecutive initial conditions to account for seasonal variability in forecast skill. The $N_f$ and $D$ values are specified in Table~\ref{tab:IC}. The appropriately colored shaded regions around the ACC curves indicate the region between the first and third quartile values of the corresponding quantity at each time step. We also plot the latitude weighted RMSE curves for the FourCastNet and IFS models in Figure~\ref{fig:acc_rmse} in Appendix~\ref{app:acc_extra_results}}
	\label{fig:IFS_AFNO}
\end{figure}

Figure~\ref{fig:IFS_AFNO}(a-f) shows the latitude weighted ACC for the FourCastNet model forecasts (Red line with markers) and the corresponding matched IFS forecasts (Blue line with markers) for the variables (a) $U_{10}$, (b) $TP$, (c) $T_{2m}$, (d) $Z_{500}$, (e) $T_{850}$, (f) $V_{10}$. The ACC and RMSE values are averaged over $N_f$ initial conditions with an interval of $D$ days between consecutive initial conditions, where the $N_f$ and $D$ values are specified in Table~\ref{tab:IC}. The shaded regions around the ACC curves indicate the region between the first and third quartile values of the corresponding quantity at each time step. The corresponding RMSE plots are shown in Figure~\ref{fig:IFS_AFNO} in Appendix~\ref{app:acc_extra_results}

In general, the FourCastNet predictions are very competitive with IFS, with our model achieving similar ACC and RMSE over a horizon of several days. At shorter lead times ($\sim48$hrs or less), we actually outperform the IFS model in ACC and/or RMSE for key variables like precipitation, winds, and temperature. Remarkably, we achieve this accuracy using only part of the full variable set available to the IFS model, and we do so at a fraction of the compute cost (see section \ref{sec:speed} for a detailed speed comparison between models). We also obtain excellent forecast accuracy on the rest of the variables predicted by our backbone model, which we include in Appendix \ref{app:acc_extra_results}.

\label{sec:afno_ifs}

\subsection{Ensemble Forecasts Using FourCastNet}
\label{sec:ensemble}

Ensemble forecasts have become a crucial component of numerical weather prediction \citep{palmer2019ecmwf}, and consume the largest share of compute costs at operational weather forecasting centers \citep{bauer2020ecmwf}. An ensemble forecast improves upon a single deterministic forecast by modeling multiple possible trajectories of a system. For a chaotic atmosphere with uncertain initial conditions, ensemble forecasting helps quantify the likelihood of extreme events and improves the accuracy of long-term predictions. Thus, rapidly generating large ensemble forecasts is an extremely promising direction for DL-based weather models \citep{weyn2021sub}, which can provide immense speedups over traditional NWP models.
 NWP models such as the IFS perform ensemble forecasting with up to 51 ensemble members. The initial conditions for the ensemble forecasts are obtained by perturbing the analysis state obtained from data assimilation. 
 
 As seen in Section~\ref{sec:hurricane}, ensemble forecasting is useful for generating probabilistic forecasts of extreme events such as hurricanes. While the individual perturbed ensemble members typically show lower forecast skill than the unperturbed `control' forecast, the mean of a large number of such perturbed ensemble members has better forecast skill than the control. 

In Section~\ref{sec:speed}, we estimate that FourCastNet is roughly 45,000 times faster than a traditional NWP model. This speed allows us to consider probabilistic ensemble forecasting with massive ensemble sizes. Ensemble weather forecasts using FourCastNet are highly computationally efficient because (1.) Inference time for a single forecast on a GPU is very fast and (2.) An ensemble of initial conditions can be folded into the the `batch' dimension in a tensor and as such, inference on a large batch ($O(100)$ or more) of initial conditions using a few GPUs is straightforward.

\begin{figure}
    \centering
    \includegraphics[width = 0.97\textwidth]{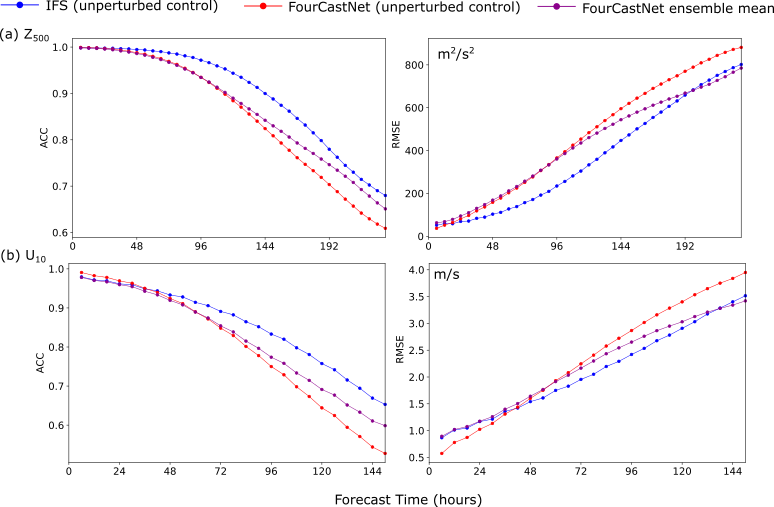}
    \caption{Illustration of the improvement in forecast skill of FourCastNet by utilizing large ensembles. We compare the forecast skill of the unperturbed `control' forecasts using FourCastNet (red) with the mean of a 100-member ensemble forecast using FourCastNet (magenta) for $Z_{500}$ (panel a) and $U_{10}$ (panel b). The IFS unperturbed control forecast is included for reference (blue). All ACC and RMSE plots are averaged over several forecasts over a year as indicated in Table~\ref{tab:IC} in Appendix~\ref{app:hyperparams} to account for seasonal and day-to-day variability in forecast skill. We find that the 100-member FourCastNet ensemble mean is more skillful than the FourCastNet control at longer forecast lead times. The 100-member FourCastNet ensemble mean shows significant improvement over the unperturbed FourCastNet control forecast beyond 70 hours for $U_{10}$ and 100 hours for $Z_{500}$. Due to the challenge of clearly disambiguating in a single plot the forecast spread arising from simulated initial condition uncertainty and the forecast spread due to seasonal and day-to-day variability, we choose not to visualize the spread in ACC and RMSE over the $N_f$ forecasts and simply report the mean.}
    \label{fig:ensemble}
\end{figure}

As a simple test of ensemble forecasting, we generate an ensemble forecast using FourCastNet from a given ERA5 initial condition by perturbing the initial condition using Gaussian random noise. This allows us to simulate initial condition uncertainty due to errors in the estimate of the starting state of the forecast. This method of ensemble generation is the same as methods used in Ensemble Kalman Filtering (EnKF)~\citep{evensen2003ensemble} for background forecast covariance estimation and not too dissimilar from the way operational NWP models generate perturbed initial conditions. Thus, given an initial condition $\vv{X}_{\text{true}}(k)$ from our out-of-sample testing dataset, we generate an ensemble of $E$ perturbed initial conditions $\lbrace \vv{X}^{(e)}(k) = \hat{\vv{X}}_{\text{true}}(k) + \sigma\xi  \rbrace_{e = 1}^E$, where $\hat{\vv{X}}_{\text{true}}(k)$ is the standardized initial condition with zero mean and unit variance and $\xi \sim \mathcal{N}(\vv{0}, \vv{1})$ is a normally distributed random variable of the same shape as $\vv{X}_{\text{true}}$ and with unit mean and variance. The perturbations are scaled by a factor $\sigma=0.3$. We refer to the forecast starting from the unperturbed initial condition as the control forecast. We generate an ensemble of perturbed forecasts each starting from a perturbed initial conditions and compute the ensemble mean of the perturbed forecasts at every forecast time step. We compute a control forecast and an ensemble mean forecast for $N_f$ initial conditions separated by $D$ days as stated in Table~\ref{tab:IC}. We report the mean ACC and RMSE over all $N_f$ initial conditions for both the control and the mean forecast in Figure~\ref{fig:ensemble}. 

Figure~\ref{fig:ensemble} shows the ACC and RMSE of the FourCastNet ensemble mean (magenta line with markers) and FourCastNet unperturbed control (red line with markers) forecasts along with the unperturbed control IFS model (blue line with markers) forecasts for reference. It is challenging to unambiguously visualize in a single plot, both the spread due to simulated initial condition uncertainty in an ensemble forecast and the spread due to seasonal and day-to-day variability. As such, we do not visualize the spread in ACC and RMSE over the $N_f$ forecasts and simply report the mean.

Indeed, in Figure~\ref{fig:ensemble},  we see that the ensemble mean from our 100-member FourCastNet ensemble results in a net improvement in ACC and RMSE at longer timescales over the unperturbed control.
We do observe a marginal degradation in skill for the ensemble mean at short ($<48$hr) lead times, as averaging over the individual ensemble members likely averages over relevant fine-scale features. Nevertheless, these ensemble forecasts are impressive, and warrant further work in how to optimally choose ensemble members. In addition to perturbing initial conditions with Gaussian noise, as we do here, it is possible and likely worthwhile to introduce more nuanced perturbations to both the initial conditions as well as the model itself. This is a promising direction of research for future work.

\subsection{Forecast Skill Over Land For Near-surface Wind Speed}

\begin{figure}
    \centering
    \includegraphics[width= \textwidth]{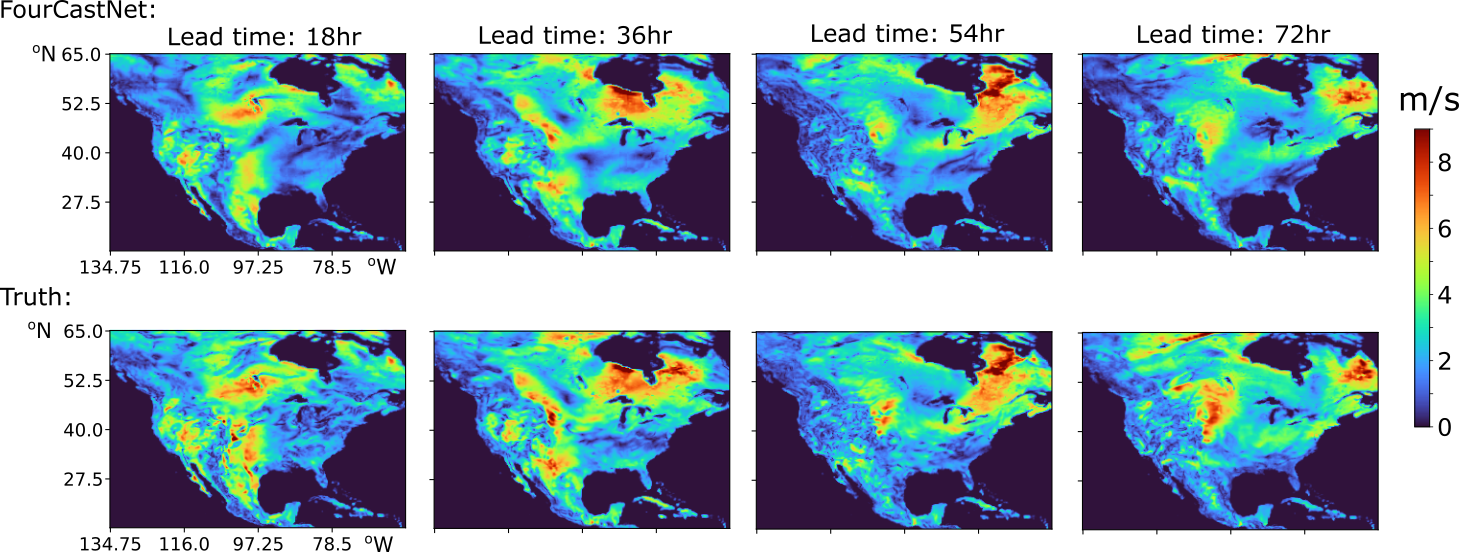}
    \caption{The FourCastNet model shows excellent skill on forecasting overland wind speed, a challenging problem due to topographic features such as mountains and lakes. This is a significant result for wind energy resource planning, as windfarms are located on land or just offshore. The figure shows The 10m wind speed ($\sqrt{U_{10}^2 + V_{10}^2}$) forecast (top four panels) generated by FourCastNet and corresponding ground truth (bottom four panels) for forecast lead times of 18 hours, 36 hours, 54 hours and 72 hours. The forecast was initialized with an initial condition at calendar time 06:00:00 on July 4 2018 UTC. To better visualize the forecast skill of our model over landmass, we plot the 10m wind speed forecast and ground truth over North America after zeroing out the fields over the ocean by multiplying the forecast and ground truth with the land masking factor $\Phi_{\mathrm{land}}$ described in Appendix \ref{app:acc_extra_results}.}
    \label{fig:overland}
\end{figure}

Most wind farms are located on land or just off of coastlines, so accurately modeling near-surface wind speed over these regions is of critical importance to wind energy resource planning. To demonstrate the accuracy of FourCastNet predictions over landmasses, we plot the 10m wind speed ($\sqrt{U_{10}^2 + V_{10}^2}$) forecast and ground truth over North America in Figure~\ref{fig:overland}. We find that FourCastNet can qualitatively capture the spatial patterns and intensities of surface winds with impressive accuracy up to several days in advance. Moreover, the visualizations emphasize the importance of running forecasts at high resolution, as the surface wind speed exhibits significant small-scale spatial variations which would be lost with a coarser grid.

We evaluate the forecast skill of our model over land versus over oceans quantitatively in Appendix \ref{app:acc_extra_results}. By computing a separate land-masked ACC and a sea-masked ACC for the surface wind velocity components, we find that the forecast quality of our model for surface wind speed over landmass is almost as good as it is over the ocean. This is significant, as surface wind speed over land is strongly affected by orographic features such as mountains, making it in general harder to forecast surface winds over land than over the oceans.

\subsection{Extremes}
\label{sec:extremes}

\begin{figure}
	\centering
	\includegraphics[width =\textwidth]{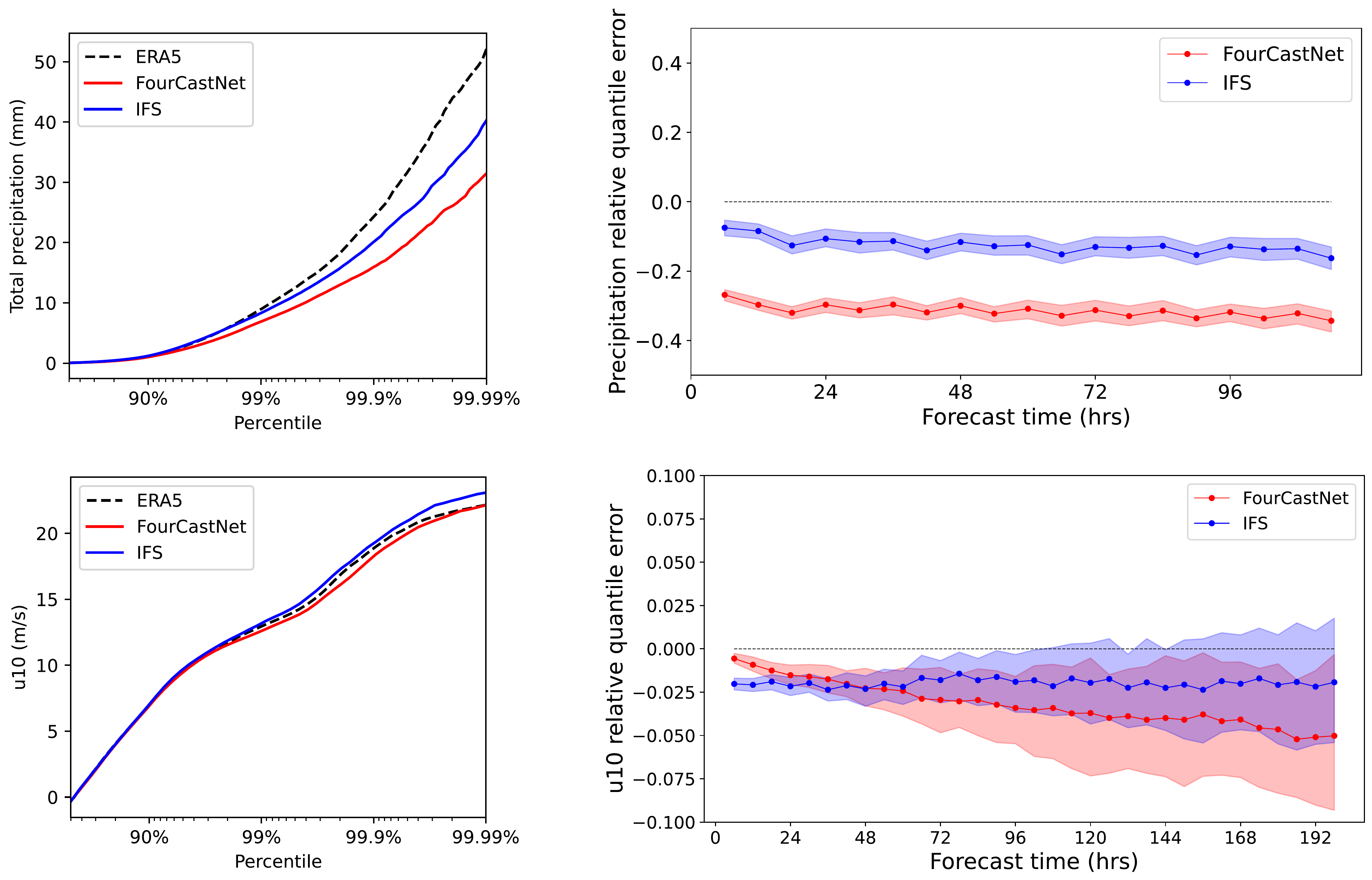}
	\caption{Comparison of extreme percentiles between ERA5, FourCastNet, and IFS. The left panel shows the top percentiles of the $TP$ and $U_{10}$ distribution at a forecast time of 24 hours, for a randomly sampled initial condition. The right panel shows the $TP$ and $U_{10}$ relative quantile error (RQE, defined in the text) as a function of forecast time, averaged over $N_f$ initial conditions in the calendar year 2018 (filled region spans the 1st and 3rd quartiles). On average, RQE trends slightly negative for both models as they under-predict the most extreme values for these variables, especially for $TP$.}
	\label{fig:extremes}
\end{figure}

We assess the ability of the FourCastNet model to capture instantaneous extremes by looking at the top quantiles of each field at a given time step. Similar to the approach in \cite{fildier21distortions}, we use 50 logarithmically-spaced quantile bins $Q = 1 - \{10^{-1}, ..., 10^{-4}\}$ (corresponding to percentiles $\{90\%, ..., 99.99\% \}$) to emphasize the most extreme values (generally, the FourCastNet predictions and ERA5 targets match closely up to around the 98$^\mathrm{th}$ percentile). We choose the 99.99$^{\mathrm{th}}$ as the top percentile  bin because percentiles beyond there sample less than 1000 pixels in each image and are subject to more variability. We show example plots of the top quantiles for $U_{10}$ and $TP$ at 24-hour forecast times in the left panel of Figure \ref{fig:extremes} (these particular forecasts were initialized at 00:00 UTC Jan 1 2018). At this particular time, both the FourCastNet and IFS models under-predict extreme precipitation, while for extreme winds in $U_{10}$ the IFS model over-predicts and FourCastNet under-predicts. To get a more comprehensive picture, we need to evaluate the model performance at multiple forecast times over multiple initial conditions in order to ascertain if there is a systematic bias in the model's predictions for extreme values.

To this end, we define the relative quantile error (RQE) at each time step $l$ as
\begin{equation}
    \mathrm{RQE} (l) = \sum_{q \in Q} (\vv{X}_{\text{pred}}^q (l)  - \vv{X}_{\text{true}}^q (l))/\vv{X}_{\text{true}}^q (l) ,
\end{equation}
where $\vv{X}^q (l)$ is the $q^{\mathrm{th}}$-quantile of $\vv{X} (l)$. RQE trends negative for a given variable if a model systematically under-predicts that variable's extremes, and we indeed find that both the FourCastNet and IFS models show a slight negative RQE over different forecast times and initial conditions for both $TP$ and $U_{10}$. This can be seen in the right-hand panel of Figure \ref{fig:extremes}. For $U_{10}$, the difference between FourCastNet and IFS is negligible and, on average, both models underestimate the extreme percentiles by just a few percentage points in RQE. 

For $TP$, the difference with respect to IFS is more pronounced, and FourCastNet underestimates the extreme percentiles by $\sim35\%$ in RQE, compared to $\sim15\%$ for IFS. This is not surprising given the forecasts visualized in Figure \ref{fig:precip}, which show the FourCastNet predictions being generally smoother than the ERA5 targets. As the extreme values tend to be concentrated in extremely small regions (sometimes down to the gridbox/pixel scale), a model that fails to fully resolve these scales will have a harder time capturing $TP$ extremes. Given the noise and uncertainties, predicting precipitation extremes is well-known to be a challenging problem, but we believe our model could be improved further by focusing more on such fine-scale features. We leave this for future work.

\section{Computational Cost of FourCastNet}\label{sec:speed}
In comparing the speed of forecast generation between FourCastNet and IFS, we have to deal with the rather difficult problem of comparing a forecast computed using a CPU cluster (in the case of the IFS model) and a forecast that is computed on a single (or perhaps a few) GPU(s) (FourCastNet). We take a nuanced approach to reporting this comparison. Our motivation is not to create a definitive apples to apples comparison and tout a single numerical factor advantage for our model, but merely to illustrate the order-of-magnitude differences in forecast generation time and also highlight the radically different perspectives of computation when comparing traditional NWP models with DL models. 
Through this comparison, we also wish to highlight the significant potential of FourCastNet and future DL models to offer an important addition to the toolkit of a meteorologist.

To estimate the forecast speed of the IFS model, we use figures provided in \citet{bauer2020ecmwf} as a baseline. In Ref.~\citep{bauer2020ecmwf}, we see that the IFS model computes a 15-day, 51-member ensemble forecast using the ``L91'' 18km resolution grid on 1530 Cray XC40 nodes with dual socket Intel Haswell processors in 82 minutes. The IFS model archived in TIGGE, which we compare the FourCastNet predictions with in Section~\ref{sec:afno_ifs}, also uses the L91 18km grid for computation (but is archived at the ERA5 resolution of 30km). Based on this information, we estimate that to compute a 24-hour 100-member ensemble forecast, the reference IFS model would require 984,000 node-seconds. We estimate the energy consumption for computing such a 100-member forecast to be 271MJ\footnote{A dual-socket Intel Haswell node draws a Thermal Design Power (TDP) of 270 Watts}.

We now estimate the latency and energy consumption of the FourCastNet model. The FourCastNet model can compute a 100-member 24-hour forecast in 7 seconds by using a single node on the Perlmutter HPC cluster which contains 4 A100 GPUs per node. This is achieved by performing batched inference on the 4 A100 GPUs using a batch size of 25. Thus, the FourCastNet model takes 7 node-seconds per forecast day for a 100-member ensemble. With a peak power consumption of 1kW per node, we estimate this 100-member 24-hour forecast to use 8kJ of energy.

We attempt to account for the resolution difference between the 18km L91 model and the 30km FourCastNet model by additionally reporting the inference time for an 18km FourCastNet model. Since we did not train the FourCastNet with 18km resolution data (due to the lack of such a publicly available dataset), the reported numbers simply estimate the computational costs for such a hypothetical model by performing inference on data and model parameters interpolated to 18km resolution from the original 30km resolution.

Table~\ref{tab:compute} provides a comparison of computational speed and energy consumption of the IFS L91 18km model and the FourCastNet model at 30km resolution, as well as the extrapolated 18km resolution. These results suggest that the FourCastNet model can compute a 100-member ensemble forecast using vastly fewer nodes, at a speed that is between 45,000 times faster (at the 18 km resolution) and 145,000 times faster (at the 30 km resolution) on a node-to-node comparison. By the same estimates, FourCastNet has an energy consumption that is between 12,000 (18 km) and 24,000 (30 km) times lower than that of the IFS model.

\begin{table}[h]
\centering
\begin{tabular}{|lcccc|}
\hline
\multicolumn{5}{|l|}{Latency and Energy consumption for a 24-hour 100-member ensemble forecast}                                                                                                                                                                                                                                \\ \hline
\multicolumn{1}{|l|}{}                                                                 & \multicolumn{1}{c|}{IFS}    & \multicolumn{1}{c|}{\begin{tabular}[c]{@{}c@{}}FCN - 30km\\ (actual)\end{tabular}} & \multicolumn{1}{c|}{\begin{tabular}[c]{@{}c@{}}FCN - 18km\\ (extrapolated)\end{tabular}} & IFS / FCN(18km) Ratio \\ \hline
\multicolumn{1}{|l|}{Nodes required}                                                   & \multicolumn{1}{c|}{3060}   & \multicolumn{1}{c|}{1}                                                             & \multicolumn{1}{c|}{2}                                                                   & \textbf{1530}           \\ \hline
\multicolumn{1}{|l|}{\begin{tabular}[c]{@{}l@{}}Latency\\ (Node-seconds)\end{tabular}} & \multicolumn{1}{c|}{984000} & \multicolumn{1}{c|}{7}                                                             & \multicolumn{1}{c|}{22}                                                                  & \textbf{44727}          \\ \hline
\multicolumn{1}{|l|}{\begin{tabular}[c]{@{}l@{}}Energy Consumed\\ (kJ)\end{tabular}}   & \multicolumn{1}{c|}{271000} & \multicolumn{1}{c|}{7}                                                             & \multicolumn{1}{c|}{22}                                                                  & \textbf{12318}          \\ \hline
\end{tabular}
\caption{The FourCastNet model can compute a 100-member ensemble forecast on a single 4GPU A100 node. In comparison, the IFS model needs 3060 nodes for such a forecast. In this table, we provide information about latency and energy consumption for the FourCastNet model in comparison with the IFS model. The FourCastNet model at a 30km resolution is about 145,000 times faster on a single-node basis than the IFS model. We can also estimate the cost of generating an 18km resolution forecast using FourCastNet. Such a hypothetical 18km model would be about 45,000 times faster than the IFS on a single-node basis. The FourCastNet model at 30km resolution uses 24,000 times less energy to compute the ensemble forecast than the IFS model, while a hypothetical FourCastNet model at 18km resolution would use 12000 times less energy.}
\label{tab:compute}
\end{table}

This comparison comes with several caveats. The IFS model generates forecasts that are provably physically consistent, while FourCastNet in its current iteration does not impose physics constraints. The IFS model also outputs an order of magnitude more variables at as many as 100 vertical levels. Notably, the IFS model is generally more accurate than FourCastNet (although in several variables, the DL model approaches the accuracy of the IFS model and exceeds it for precipitation in certain cases). On the other hand, it is worth noting our rudimentary speed assessments of FourCastNet do not employ any of the common optimizations used for inference of DL models (e.g., model distillation, pruning, quantization, or reduced precision). We expect implementing these would greatly accelerate our speed of inference, and lead to further gains in computational efficiency over IFS.

While the above caveats are important, it is fair to say that if one were only interested in limited-purpose forecasting (e.g., a wind farm operator interested in short-term surface wind speed forecasts), FourCastNet would be a very attractive option as the infrastructure requirements are minimal. FourCastNet can generate a 10-day, global forecast at full ERA5 resolution using a single device, which is simply not possible with IFS, and such a forecast completes in seconds. This means one could generate reasonably accurate forecasts using a tabletop computer with a single GPU, rather than needing a substantial portion of a compute cluster. Similarly, only a handful of GPUs are needed for generating ensemble forecasts with 100s of ensemble members, and such ensembles run quickly and efficiently using batched inference. This greatly lowers the barrier to entry for doing data-assimilation and uncertainty quantification, and future work in this direction is warranted to explore these possibilities.

\section{Comparison Against State-of-the-art DL Weather Prediction}
To the best of our knowledge, the current state-of-the-art DL weather prediction model is the DLWP model of \citet{weyn2020improving}---they employ a deep convolutional network with a cubed-sphere remapped coordinate system to predict important weather forecast variables. The authors work with a coarser resolution of $2^{\circ}$ and forecast variables relating to geopotential heights, geopotential thickness, and 2-m temperature (see \citep{weyn2020improving} for further details). The FourCastNet model predicts more variables than the DLWP model at a resolution that is higher than the DLWP model by a factor of 8. The significantly higher resolution of the FourCastNet model resolves small-scale features present in variables such as wind velocities and precipitation allowing us to resolve important phenomena such as hurricanes, extreme precipitation and atmospheric rivers. This would not be possible at a lower resolution such as $2^{\degree}$ (and almost entirely a futile exercise at a $5^{\degree}$ resolution.) For reference, we have visualized the MSLP over the trajectory of hurricane Michael at a resolution of $2^{\degree}$ in Figure~\ref{fig:lowres_mslp} of Appendix \ref{app:MSLP_hurricane}. Thus, the FourCastNet model has many characteristics that make it superior to the prior SOTA DLWP model. Nonetheless, we undertake a comparison of forecasts generated by the FourCastNet model with those of the DLWP model by coarsening the FourCastNet outputs to bring them to a resolution comparable to that of the DLWP model. We emphasize that this comparison has been provided only for the sake of completeness. Coarsening our forecasts and making them less effective in order to accommodate a prior benchmark at a lower resolution is not fair to our model.

\begin{figure}
    \centering
    \includegraphics[width = 0.9\textwidth]{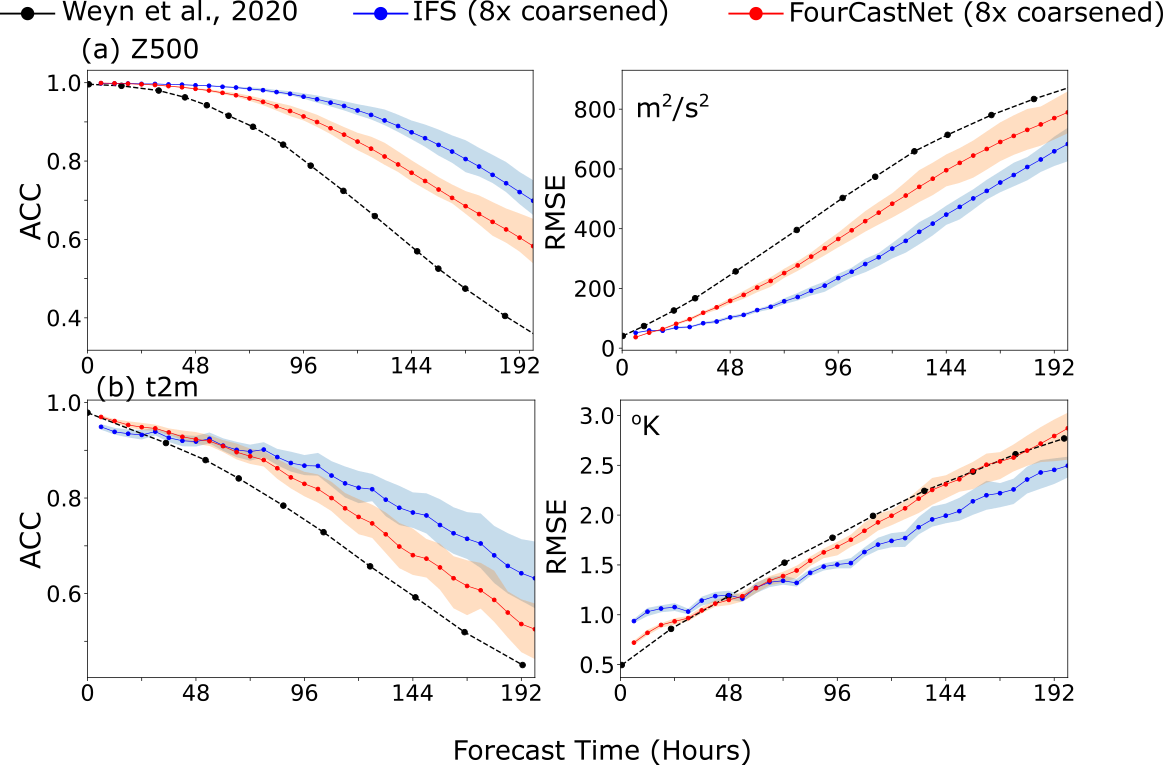}
    \caption{Comparison of ACC and RMSE metrics between the (downsampled) FourCastNet predictions, (downsampled) IFS, and baseline state-of-the-art DLWP model \citep{weyn2020improving} for (a) $Z_{500}$ and (b) $T_{2m}$. We observe that the FourCastNet predictions show significant improvement over the baseline model. We also note that the FourCastNet generates predictions that have a higher resolution by a factor of 8, and is thus able to resolve many more important small-scale features than the DLWP model.}
    \label{fig:baseline}
\end{figure}

 We downsample our predictions eight times (in each direction, using bilinear interpolation) to coarsen them to a resolution that is comparable to that of the DLWP model. Since the two variables reported in the DLWP results are $Z_{500}$ and $T_{2m}$, we re-compute our ACC and RMSE metrics for those two variables. We also note that the ACC metric in the DLWP baseline was computed using daily climatology (we use a time-averaged climatology in this work, motivated by \citep{rasp2020weatherbench}) and, hence, we modify our ACC computation using the same definition for a fair comparison. We show our comparisons for ACC and RMSE in Figure~\ref{fig:baseline}. We observe that even at the lower resolution of the DLWP work, the FourCastNet model predictions show significant improvement over the current state-of-the-art DLWP model in both variables. Additionally, the FourCastNet model operates at a resolution that is 8 times higher than the DLWP model allowing it to resolve many important small-scale phenomena.

\section{Implications, Discussion, and Future Work}

FourCastNet is a novel global data-driven DL-based weather forecasting model based on the FNO and AFNO \citep{li2021fourier, guibas2021adaptive}. FourCastNet's speed, computational cost, energy footprint, and capacity for generating large ensembles has several important implications for science and society. In particular, FourCastNet's high-resolution, high-fidelity wind and precipitation forecasts are of tremendous value. Even though FourCastNet was developed in less than a year and has only a fraction of the number of variables and vertical levels compared to NWP, its accuracy is comparable to the IFS model and better than state-of-the-art DL weather prediction models \citep{weyn2021sub, rasp2020weatherbench} on short timescales. We anticipate that with additional resources and further development, FourCastNet could match the capabilities of current NWP models on all timescales and at all vertical levels of the atmosphere. 

\subsection{Implications}

FourCastNet's predictions are four to five orders of magnitude faster than traditional NWP models. This has two important implications. First, large ensembles of thousands of members can be generated in seconds, thus enabling estimation of well-calibrated and constrained uncertainties in extremes with higher confidence than current NWP ensembles that have at most approximately 50 members owing to their high computational cost. Fast generation of 1,000-member ensembles dramatically changes what is possible in probabilistic weather forecasting, including improving reliability of early warnings of extreme weather events and enabling rapid assessment of their impacts. Second, FourCastNet is suitable for rapidly testing hypotheses about mechanisms of weather variability and their predictability. 

The unprecedented accuracy in short-range forecasts of precipitation and its extremes has potentially massive benefits for society such as enabling rapid responses for disaster mitigation. Furthermore, a highly accurate DL-based \textit{diagnostic} precipitation model provides the flexibility to input prognostic variables from different models or observational sources. 

For the wind energy industry, FourCastNet's rapid and reliable high-resolution wind forecasts can help mitigate disasters from extreme wind events and enables planning for fluctuations in wind power output. Wind farm designers can benefit from fast and reliable high-resolution wind forecasts to optimize wind farm layouts that account for a wide variety of wind and weather conditions.

\subsection{Discussion}

FourCastNet's skill improves with increasing number of modeled variables. A larger model trained on more variables, perhaps even on entire 3D atmospheric fields, may extend prediction horizons even further and with better uncertainty estimates. Not far in the future, FourCastNet could be trained on all nine petabytes of the ERA5 dataset to predict all currently predicted variables in NWP at all atmospheric levels. Although the cost of training such a model will be huge, fast inference will enable rapid predictions of entire 3D fields in a few seconds. Such an advancement will likely revolutionize weather prediction. 

Due to the current absence of a data-assimilation component, FourCastNet cannot yet generate up-to-the-minute weather forecasts. If observations are available, however, such a component could be readily added given the ease of generating large ensembles for methods such as Ensemble Kalman Filtering with data-driven background covariance estimation~\citep{chattopadhyay2020deep}. Therefore, in principle, future iterations of FourCastNet could be trained on observational data. This will enable real-time weather prediction by initializing the model with real-time observations. 

With ever-increasing demands for very high-resolution forecasts, NWP has seen a steady growth in resolution. The increase in computational cost of NWP for a doubling of resolution is nearly 12-fold ($2^{3.5}$). Current IFS forecasts are at 9-km resolution but we require forecasts at sub-km resolution for improvements in a wide variety of applications, such as energy and agricultural planning, transportation, and disaster mitigation. Simultaneously, DL continues to have ever-increasing accuracy and predictive power with larger models that have hundreds of billions of parameters \citep{rajbhandari2020zero}. With advances in large-scale DL we expect that FourCastNet can be trained to predict weather on sub-km scales. Even though training such a large DL model will be computationally expensive, since inference of large DL models can still be done rapidly \citep{deepspeed}, a sub-km resolution version of FourCastNet will have even more dramatic speedup over sub-km resolution NWP, likely more than six orders of magnitude. 

DLWP has shown good skill on S2S timescales \citep{weyn2021sub}. FourCastNet has better skill at short timescales (up to two weeks). We envision a coupled model using a two-timescale approach that combines DLWP and FourCastNet with two-way interactions to achieve unprecedented accuracies on short-, medium-, and long-range weather forecasts. 

FourCastNet is a purely data-driven DL weather model. The physical systems of weather and climate are governed by the laws of nature, some of which are well-understood, such as Navier-Stokes equations for the fluid dynamics of atmosphere and oceans. Weather forecasts and climate predictions that obey known physical laws are more trustworthy than those that do not. Furthermore, models that obey the laws of physics are more likely to be robust under climate change. An emerging field in AI applications in the sciences is \textit{Physics-informed Machine Learning} \citep{kashinath2021physics}. The Fourier Neural Operator has been extended to be physics-informed \citep{PINO}. Future versions of FourCastNet will incorporate physical laws. A physics-informed version of FourCastNet could be trained with fewer datapoints. This benefit is particularly valuable at higher resolutions in order to reduce the data volume requirements for training. FourCastNet could also combine with a physics-based NWP model \citet{arcomano2021alpha}, to generate long-term stable forecasts over S2S timescales.

An important question that remains unanswered is whether FourCastNet generalizes under climate change. FourCastNet was trained on data from 1979 to 2015 and tested on data from 2016 to 2020. We know that Earth's climate has changed over this period of time. Therefore, FourCastNet has been trained on data from a changing climate. However, FourCastNet may not predict weather reliably under extreme climate change expected in the decades to come. A future version will initialize with climate model output to evaluate FourCastNet's performance under different warming scenarios. A grand challenge for the climate community is to predict the changing behavior of extreme weather events under climate change, such as their frequency, intensity, and spatio-temporal nature. Once FourCastNet achieves high fidelity under extreme climate change, it can address this grand challenge. 

\section*{Acknowledgements}
We would like to acknowledge helpful comments and suggestions by Peter Dueben from ECMWF. We thank the researchers at ECMWF for their open data sharing and maintaining the ERA5 dataset without which this work would not have been possible. This research used resources of the National Energy Research Scientific Computing Center (NERSC), a U.S. Department of Energy Office of Science User Facility located at Lawrence Berkeley National Laboratory, operated under Contract No. DE-AC02-05CH11231. A.K. and P. Hassanzadeh were partially supported by ONR grant N00014-20-1-2722. We thank the staff and administrators of the Perlmutter computing cluster at NERSC, NVIDIA Selene computing cluster administrators, Atos, and J{\"u}lich Supercomputing Center for providing computing support. JP and KK would like to thank Sanjay Choudhry and the NVIDIA Modulus team for their support. JP, SS, P. Harrington and KK would like to thank Wahid Bhimji for helpful comments.

\bibliographystyle{unsrtnat}

\begin{thebibliography}{38}
\providecommand{\natexlab}[1]{#1}
\providecommand{\url}[1]{\texttt{#1}}
\expandafter\ifx\csname urlstyle\endcsname\relax
  \providecommand{\doi}[1]{doi: #1}\else
  \providecommand{\doi}{doi: \begingroup \urlstyle{rm}\Url}\fi

\bibitem[Bauer et~al.(2015)Bauer, Thorpe, and Brunet]{bauer2015quiet}
Peter Bauer, Alan Thorpe, and Gilbert Brunet.
\newblock The quiet revolution of numerical weather prediction.
\newblock \emph{Nature}, 525\penalty0 (7567):\penalty0 47--55, 2015.

\bibitem[Alley et~al.(2019)Alley, Emanuel, and Zhang]{alley2019advances}
Richard~B Alley, Kerry~A Emanuel, and Fuqing Zhang.
\newblock Advances in weather prediction.
\newblock \emph{Science}, 363\penalty0 (6425):\penalty0 342--344, 2019.

\bibitem[Richardson(2007)]{richardson2007weather}
Lewis~Fry Richardson.
\newblock \emph{Weather prediction by numerical process}.
\newblock Cambridge university press, 2007.

\bibitem[Schultz et~al.(2021)Schultz, Betancourt, Gong, Kleinert, Langguth,
  Leufen, Mozaffari, and Stadtler]{schultz2021can}
MG~Schultz, C~Betancourt, B~Gong, F~Kleinert, M~Langguth, LH~Leufen, Amirpasha
  Mozaffari, and S~Stadtler.
\newblock Can deep learning beat numerical weather prediction?
\newblock \emph{Philosophical Transactions of the Royal Society A},
  379\penalty0 (2194):\penalty0 20200097, 2021.

\bibitem[Balaji(2021)]{balaji2021climbing}
V~Balaji.
\newblock Climbing down charney’s ladder: machine learning and the
  post-dennard era of computational climate science.
\newblock \emph{Philosophical Transactions of the Royal Society A},
  379\penalty0 (2194):\penalty0 20200085, 2021.

\bibitem[Irrgang et~al.(2021)Irrgang, Boers, Sonnewald, Barnes, Kadow, Staneva,
  and Saynisch-Wagner]{irrgang2021towards}
Christopher Irrgang, Niklas Boers, Maike Sonnewald, Elizabeth~A Barnes,
  Christopher Kadow, Joanna Staneva, and Jan Saynisch-Wagner.
\newblock Towards neural {E}arth system modelling by integrating artificial
  intelligence in {E}arth system science.
\newblock \emph{Nature Machine Intelligence}, 3\penalty0 (8):\penalty0
  667--674, 2021.

\bibitem[Reichstein et~al.(2019)Reichstein, Camps-Valls, Stevens, Jung,
  Denzler, Carvalhais, et~al.]{reichstein2019deep}
Markus Reichstein, Gustau Camps-Valls, Bjorn Stevens, Martin Jung, Joachim
  Denzler, Nuno Carvalhais, et~al.
\newblock Deep learning and process understanding for data-driven earth system
  science.
\newblock \emph{Nature}, 566\penalty0 (7743):\penalty0 195--204, 2019.

\bibitem[Scher and Messori(2018)]{scher2018predicting}
Sebastian Scher and Gabriele Messori.
\newblock Predicting weather forecast uncertainty with machine learning.
\newblock \emph{Quarterly Journal of the Royal Meteorological Society},
  144\penalty0 (717):\penalty0 2830--2841, 2018.

\bibitem[Scher and Messori(2019)]{scher2019weather}
Sebastian Scher and Gabriele Messori.
\newblock Weather and climate forecasting with neural networks: using general
  circulation models (gcms) with different complexity as a study ground.
\newblock \emph{Geoscientific Model Development}, 12\penalty0 (7):\penalty0
  2797--2809, 2019.

\bibitem[Chattopadhyay et~al.(2020{\natexlab{a}})Chattopadhyay, Nabizadeh, and
  Hassanzadeh]{chattopadhyay2019analog}
Ashesh Chattopadhyay, Ebrahim Nabizadeh, and Pedram Hassanzadeh.
\newblock Analog forecasting of extreme-causing weather patterns using deep
  learning.
\newblock \emph{Journal of Advances in Modeling Earth Systems}, 12\penalty0
  (2):\penalty0 e2019MS001958, 2020{\natexlab{a}}.

\bibitem[Weyn et~al.(2019)Weyn, Durran, and Caruana]{weyn2019can}
Jonathan~A Weyn, Dale~R Durran, and Rich Caruana.
\newblock Can machines learn to predict weather? using deep learning to predict
  gridded 500-hpa geopotential height from historical weather data.
\newblock \emph{Journal of Advances in Modeling Earth Systems}, 11\penalty0
  (8):\penalty0 2680--2693, 2019.

\bibitem[Weyn et~al.(2020)Weyn, Durran, and Caruana]{weyn2020improving}
Jonathan~A Weyn, Dale~R Durran, and Rich Caruana.
\newblock Improving data-driven global weather prediction using deep
  convolutional neural networks on a cubed sphere.
\newblock \emph{Journal of Advances in Modeling Earth Systems}, 12\penalty0
  (9):\penalty0 e2020MS002109, 2020.

\bibitem[Weyn et~al.(2021)Weyn, Durran, Caruana, and
  Cresswell-Clay]{weyn2021sub}
Jonathan~A Weyn, Dale~R Durran, Rich Caruana, and Nathaniel Cresswell-Clay.
\newblock Sub-seasonal forecasting with a large ensemble of deep-learning
  weather prediction models.
\newblock \emph{arXiv preprint arXiv:2102.05107}, 2021.

\bibitem[Rasp et~al.(2020)Rasp, Dueben, Scher, Weyn, Mouatadid, and
  Thuerey]{rasp2020weatherbench}
Stephan Rasp, Peter~D Dueben, Sebastian Scher, Jonathan~A Weyn, Soukayna
  Mouatadid, and Nils Thuerey.
\newblock Weatherbench: a benchmark data set for data-driven weather
  forecasting.
\newblock \emph{Journal of Advances in Modeling Earth Systems}, 12\penalty0
  (11):\penalty0 e2020MS002203, 2020.

\bibitem[Rasp and Thuerey(2021{\natexlab{a}})]{rasp2021data}
Stephan Rasp and Nils Thuerey.
\newblock Data-driven medium-range weather prediction with a resnet pretrained
  on climate simulations: A new model for weatherbench.
\newblock \emph{Journal of Advances in Modeling Earth Systems}, 13\penalty0
  (2):\penalty0 e2020MS002405, 2021{\natexlab{a}}.

\bibitem[Rasp and Thuerey(2020)]{rasp2020purely}
Stephan Rasp and Nils Thuerey.
\newblock Purely data-driven medium-range weather forecasting achieves
  comparable skill to physical models at similar resolution.
\newblock \emph{arXiv preprint arXiv:2008.08626}, 2020.

\bibitem[Chattopadhyay et~al.(2021)Chattopadhyay, Mustafa, Hassanzadeh, Bach,
  and Kashinath]{chattopadhyay2021towards}
Ashesh Chattopadhyay, Mustafa Mustafa, Pedram Hassanzadeh, Eviatar Bach, and
  Karthik Kashinath.
\newblock Towards physically consistent data-driven weather forecasting:
  Integrating data assimilation with equivariance-preserving spatial
  transformers in a case study with era5.
\newblock \emph{Geoscientific Model Development Discussions}, pages 1--23,
  2021.

\bibitem[Arcomano et~al.(2020)Arcomano, Szunyogh, Pathak, Wikner, Hunt, and
  Ott]{arcomano2020machine}
Troy Arcomano, Istvan Szunyogh, Jaideep Pathak, Alexander Wikner, Brian~R Hunt,
  and Edward Ott.
\newblock A machine learning-based global atmospheric forecast model.
\newblock \emph{Geophysical Research Letters}, 47\penalty0 (9):\penalty0
  e2020GL087776, 2020.

\bibitem[Chantry et~al.(2021)Chantry, Christensen, Dueben, and
  Palmer]{chantry2021opportunities}
Matthew Chantry, Hannah Christensen, Peter Dueben, and Tim Palmer.
\newblock Opportunities and challenges for machine learning in weather and
  climate modelling: hard, medium and soft ai.
\newblock \emph{Philosophical Transactions of the Royal Society A},
  379\penalty0 (2194):\penalty0 20200083, 2021.

\bibitem[Gr{\"o}nquist et~al.(2021)Gr{\"o}nquist, Yao, Ben-Nun, Dryden, Dueben,
  Li, and Hoefler]{gronquist2021deep}
Peter Gr{\"o}nquist, Chengyuan Yao, Tal Ben-Nun, Nikoli Dryden, Peter Dueben,
  Shigang Li, and Torsten Hoefler.
\newblock Deep learning for post-processing ensemble weather forecasts.
\newblock \emph{Philosophical Transactions of the Royal Society A},
  379\penalty0 (2194):\penalty0 20200092, 2021.

\bibitem[Rasp and Thuerey(2021{\natexlab{b}})]{rasp_2020_resnet}
Stephan Rasp and Nils Thuerey.
\newblock Data-driven medium-range weather prediction with a resnet pretrained
  on climate simulations: A new model for weatherbench.
\newblock \emph{Journal of Advances in Modeling Earth Systems}, page
  e2020MS002405, 2021{\natexlab{b}}.

\bibitem[Guibas et~al.(2022)Guibas, Mardani, Li, Tao, Anandkumar, and
  Catanzaro]{guibas2021adaptive}
John Guibas, Morteza Mardani, Zongyi Li, Andrew Tao, Anima Anandkumar, and
  Bryan Catanzaro.
\newblock Adaptive {Fourier Neural Operators}: Efficient token mixers for
  transformers.
\newblock \emph{International Conference on Representation Learning (to
  appear)}, April 2022.

\bibitem[Dosovitskiy et~al.(2021)Dosovitskiy, Beyer, Kolesnikov, Weissenborn,
  Zhai, Unterthiner, Dehghani, Minderer, Heigold, Gelly, Uszkoreit, and
  Houlsby]{dosovitskiy2021image}
Alexey Dosovitskiy, Lucas Beyer, Alexander Kolesnikov, Dirk Weissenborn,
  Xiaohua Zhai, Thomas Unterthiner, Mostafa Dehghani, Matthias Minderer, Georg
  Heigold, Sylvain Gelly, Jakob Uszkoreit, and Neil Houlsby.
\newblock An image is worth 16x16 words: Transformers for image recognition at
  scale, 2021.

\bibitem[Li et~al.(2021{\natexlab{a}})Li, Kovachki, Azizzadenesheli, Liu,
  Bhattacharya, Stuart, and Anandkumar]{li2021fourier}
Zongyi Li, Nikola Kovachki, Kamyar Azizzadenesheli, Burigede Liu, Kaushik
  Bhattacharya, Andrew Stuart, and Anima Anandkumar.
\newblock Fourier neural operator for parametric partial differential
  equations.
\newblock In \emph{International Conference on Learning Representations
  (ICLR)}, 2021{\natexlab{a}}.

\bibitem[Hersbach et~al.(2020)Hersbach, Bell, Berrisford, Hirahara,
  Hor{\'a}nyi, Mu{\~n}oz-Sabater, Nicolas, Peubey, Radu, Schepers, Simmons,
  Soci, Abdalla, Abellan, Balsamo, Bechtold, Biavati, Bidlot, Bonavita, Chiara,
  Dahlgren, Dee, Diamantakis, Dragani, Flemming, Forbes, Fuentes, Geer,
  Haimberger, Healy, Hogan, H{\'o}lm, Janiskov{\'a}, Keeley, Laloyaux, Lopez,
  Lupu, Radnoti, de~Rosnay, Rozum, Vamborg, Villaume, and
  Th{\'e}paut]{hersbach2020era5}
Hans Hersbach, Bill Bell, Paul Berrisford, Shoji Hirahara, Andr{\'a}s
  Hor{\'a}nyi, Joaqu{\'i}n Mu{\~n}oz-Sabater, Julien Nicolas, Carole Peubey,
  Raluca Radu, Dinand Schepers, Adrian Simmons, Cornel Soci, Saleh Abdalla,
  Xavier Abellan, Gianpaolo Balsamo, Peter Bechtold, Gionata Biavati, Jean
  Bidlot, Massimo Bonavita, Giovanna~De Chiara, Per Dahlgren, Dick Dee, Michail
  Diamantakis, Rossana Dragani, Johannes Flemming, Richard Forbes, Manuel
  Fuentes, Alan Geer, Leo Haimberger, Sean Healy, Robin~J. Hogan, El{\'i}as
  H{\'o}lm, Marta Janiskov{\'a}, Sarah Keeley, Patrick Laloyaux, Philippe
  Lopez, Cristina Lupu, Gabor Radnoti, Patricia de~Rosnay, Iryna Rozum, Freja
  Vamborg, Sebastien Villaume, and Jean-No{\"e}l Th{\'e}paut.
\newblock The {{ERA5}} global reanalysis.
\newblock \emph{Quarterly Journal of the Royal Meteorological Society},
  146\penalty0 (730):\penalty0 1999--2049, 2020.
\newblock ISSN 1477-870X.

\bibitem[Kalnay et~al.(1996)Kalnay, Kanamitsu, Kistler, Collins, Deaven,
  Gandin, Iredell, Saha, White, Woollen, et~al.]{kalnay1996ncep}
Eugenia Kalnay, Masao Kanamitsu, Robert Kistler, William Collins, Dennis
  Deaven, Lev Gandin, Mark Iredell, Suranjana Saha, Glenn White, John Woollen,
  et~al.
\newblock The ncep/ncar 40-year reanalysis project.
\newblock \emph{Bulletin of the American meteorological Society}, 77\penalty0
  (3):\penalty0 437--472, 1996.

\bibitem[Kalnay(2003)]{kalnay2003atmospheric}
Eugenia Kalnay.
\newblock \emph{Atmospheric modeling, data assimilation and predictability}.
\newblock Cambridge University Press, 2003.

\bibitem[Beven~II et~al.(2019)Beven~II, Berg, and Hagen]{hcmichael}
J.L. Beven~II, R.~Berg, and A.~Hagen.
\newblock Tropical cyclone report hurricane michael, April 2019.

\bibitem[Palmer(2019)]{palmer2019ecmwf}
Tim Palmer.
\newblock The ecmwf ensemble prediction system: Looking back (more than) 25
  years and projecting forward 25 years.
\newblock \emph{Quarterly Journal of the Royal Meteorological Society},
  145:\penalty0 12--24, 2019.

\bibitem[Bauer et~al.(2020)Bauer, Quintino, Wedi, Bonanni, Chrust, Deconinck,
  Diamantakis, D{\"u}ben, English, Flemming, et~al.]{bauer2020ecmwf}
Peter Bauer, Tiago Quintino, Nils Wedi, Antonio Bonanni, Marcin Chrust, Willem
  Deconinck, Michail Diamantakis, Peter D{\"u}ben, Stephen English, Johannes
  Flemming, et~al.
\newblock \emph{The ecmwf scalability programme: Progress and plans}.
\newblock European Centre for Medium Range Weather Forecasts, 2020.
\newblock \doi{10.21957/gdit22ulm}.
\newblock URL \url{https://www.ecmwf.int/node/19380}.

\bibitem[Evensen(2003)]{evensen2003ensemble}
Geir Evensen.
\newblock The ensemble kalman filter: Theoretical formulation and practical
  implementation.
\newblock \emph{Ocean dynamics}, 53\penalty0 (4):\penalty0 343--367, 2003.

\bibitem[Fildier et~al.(2021)Fildier, Collins, and
  Muller]{fildier21distortions}
Benjamin Fildier, William~D. Collins, and Caroline Muller.
\newblock Distortions of the rain distribution with warming, with and without
  self-aggregation.
\newblock \emph{Journal of Advances in Modeling Earth Systems}, 13\penalty0
  (2):\penalty0 e2020MS002256, 2021.
\newblock \doi{https://doi.org/10.1029/2020MS002256}.
\newblock URL
  \url{https://agupubs.onlinelibrary.wiley.com/doi/abs/10.1029/2020MS002256}.
\newblock e2020MS002256 2020MS002256.

\bibitem[Chattopadhyay et~al.(2020{\natexlab{b}})Chattopadhyay, Mustafa,
  Hassanzadeh, and Kashinath]{chattopadhyay2020deep}
Ashesh Chattopadhyay, Mustafa Mustafa, Pedram Hassanzadeh, and Karthik
  Kashinath.
\newblock Deep spatial transformers for autoregressive data-driven forecasting
  of geophysical turbulence.
\newblock In \emph{Proceedings of the 10th International Conference on Climate
  Informatics}, pages 106--112, 2020{\natexlab{b}}.

\bibitem[Rajbhandari et~al.(2020)Rajbhandari, Rasley, Ruwase, and
  He]{rajbhandari2020zero}
Samyam Rajbhandari, Jeff Rasley, Olatunji Ruwase, and Yuxiong He.
\newblock Zero: Memory optimizations toward training trillion parameter models.
\newblock In \emph{SC20: International Conference for High Performance
  Computing, Networking, Storage and Analysis}, pages 1--16. IEEE, 2020.

\bibitem[dee(2021)]{deepspeed}
Deep{S}peed: Accelerating large-scale model inference and training via system
  optimizations and compression, 2021.
\newblock URL
  \url{https://www.microsoft.com/en-us/research/blog/deepspeed-accelerating-large-scale-model-inference-and-training-via-system-optimizatio/ns-and-compression/}.

\bibitem[Kashinath et~al.(2021)Kashinath, Mustafa, Albert, Wu, Jiang,
  Esmaeilzadeh, Azizzadenesheli, Wang, Chattopadhyay, Singh,
  et~al.]{kashinath2021physics}
K~Kashinath, M~Mustafa, A~Albert, JL~Wu, C~Jiang, S~Esmaeilzadeh,
  K~Azizzadenesheli, R~Wang, A~Chattopadhyay, A~Singh, et~al.
\newblock Physics-informed machine learning: case studies for weather and
  climate modelling.
\newblock \emph{Philosophical Transactions of the Royal Society A},
  379\penalty0 (2194):\penalty0 20200093, 2021.

\bibitem[Li et~al.(2021{\natexlab{b}})Li, Zheng, Kovachki, Jin, Chen, Liu,
  Azizzadenesheli, and Anandkumar]{PINO}
Zongyi Li, Hongkai Zheng, Nikola Kovachki, David Jin, Haoxuan Chen, Burigede
  Liu, Kamyar Azizzadenesheli, and Anima Anandkumar.
\newblock Physics-informed neural operator for learning partial differential
  equations, 2021{\natexlab{b}}.

\bibitem[Arcomano et~al.(2021)Arcomano, Szunyogh, Wikner, Pathak, Hunt, and
  Ott]{arcomano2021alpha}
Troy Arcomano, Istvan Szunyogh, Alexander Wikner, Jaideep Pathak, Brian~R Hunt,
  and Edward Ott.
\newblock A hybrid approach to atmospheric modeling that combines machine
  learning with a physics-based numerical model.
\newblock \emph{Journal of Advances in Modeling Earth Systems}, 2021.

\end{thebibliography}

\clearpage

\section*{Appendix A: Model Hyperparameters}
\customlabel{app:hyperparams}{A}

We list the hyperparameters of our FourCastNet AFNO model in Table \ref{tab:hyperparameters}. In addition, we list the number of initial conditions $N_f$ and assumed temporal de-correlation time $D$ used to compute metrics in Table \ref{tab:IC}.

\begin{table}[h]
    \centering
    \begin{tabular}{l|c}
     Hyperparameter     &  Value \\
     \hline
     Global batch size & 64 \\
     Learning rate (pre-training $l_1$/fine-tuning $l_2$/$TP$ model $l_3$) & \num{5E-4}/\num{1E-4}/\num{2.5E-4} \\
     Learning rate schedule & Cosine \\
     Patch size $p \times p$ & 8 $\times$ 8 \\
     Sparsity threshold $\lambda$ & \num{1E-2} \\
     Number of AFNO blocks $n_b$ & 8 \\
     Depth & 12 \\
     MLP ratio & 4 \\
     AFNO embedding dimension & 768 \\
     Activation function & GELU \\
     Dropout & 0 \\
     \hline
    \end{tabular}
    \caption{Hyperparameters used in FourCastNet model and training. We refer to \citep{guibas2021adaptive} for further details regarding the definition of AFNO backbone model parameters.}
    \label{tab:hyperparameters}
\end{table}

\begin{table}[h]
    \centering
    \begin{tabular}{c|c|c}
        Variable & $N_f$ & $D$ (days)  \\
        \hline
        $Z_{500}$ & 36 & 9 \\
        $T_{850}$ & 36 & 9 \\
        $T_{2m}$ & 40 & 9 \\
        $U_{10}$ & 178 & 2 \\
        $V_{10}$ & 178 & 2 \\
        $TP$ & 180 & 2 \\
        \hline
    \end{tabular}
   \caption{Number of initial conditions used for computing ACC and RMSE plots with the assumed temporal de-correlation time for the variables $Z_{500}$, $T_{850}$, $T_{2m}$, $U_{10}$, $V_{10}$, $TP$. }
    \label{tab:IC}
\end{table}

\clearpage
\section*{Appendix B: MSLP visualization for Hurricane Michael}
\customlabel{app:MSLP_hurricane}{B}

\begin{figure}[ht]
  \centering
  \includegraphics[width=\linewidth]{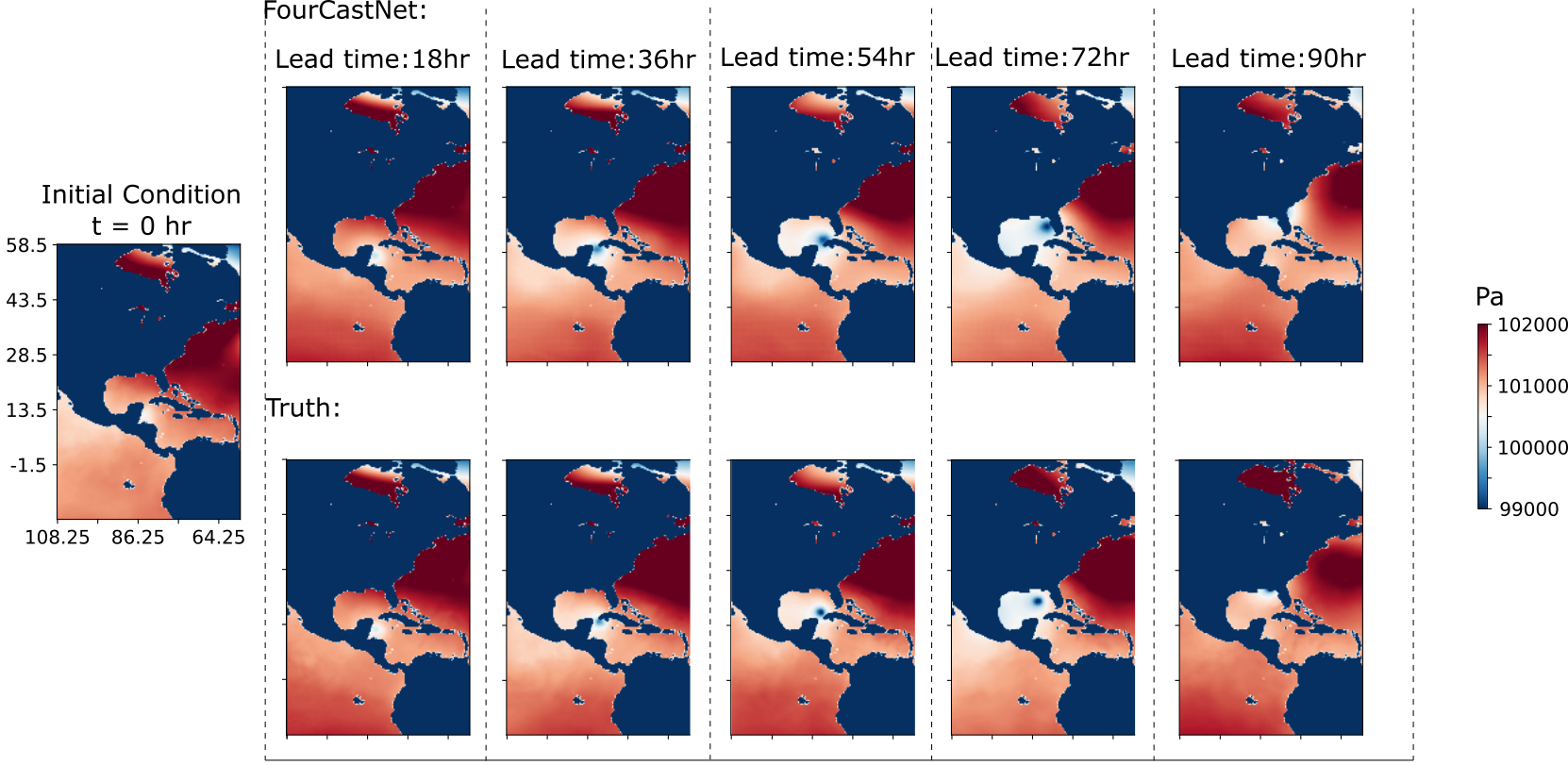}
  \caption{FourCastNet Predictions of the Mean Sea-Level Pressure with the corresponding ground truth at Forecast lead times of up to 72 hours. The forecast was initialized at a calendar time of October 7, 2018 00:00 UTC. A land-sea mask was applied to make the MSLP zero over landmass for better visualization.} 
  \label{fig:hcm_mslp}
\end{figure}

In Figure \ref{fig:hcm_mslp}, we show our model predictions for $MSLP$ in Hurricane Michael, compared to the target ERA5 snapshots at the same time-steps.  The forecast shows Hurricane Michael intensifying from a tropical depression to a hurricane as it moves towards the coast of Florida.

\begin{figure}[h]
    \centering
    \includegraphics[width = \textwidth]{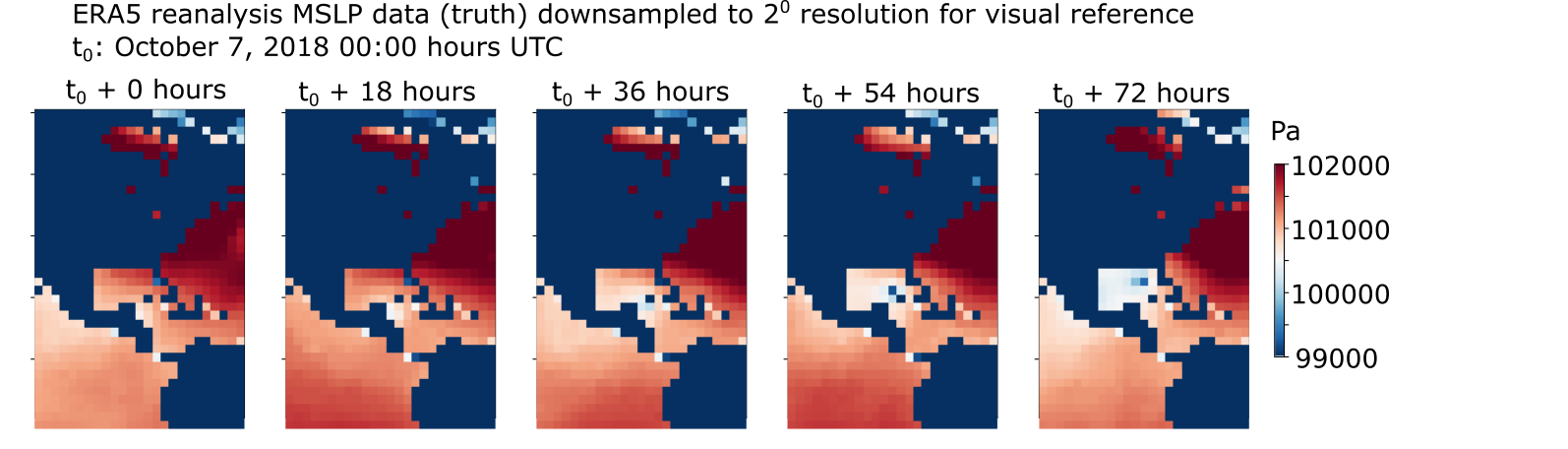}
    \caption{Mean Sea Level Pressure (MSLP) ERA5 ground truth plotted after downsampling by a factor of 8 (to a resolution of $2^{\degree}$) for visual reference. This figure highlights the importance of training a DL model at a high resolution to capture small scale phenomena such as hurricanes. At a resolution of $2^{\degree}$, small scale phenomena are not captured very well in the training data. While we did not attempt to train a model at this coarse resolution, it is reasonable to expect significantly worse performance on forecasting small scale phenomena from a model trained at this coarse resolution. }
    \label{fig:lowres_mslp}
\end{figure}

For reference, we also visualize the same ERA5 targets at a coarse resolution of $2^{\degree}$ lat-long in Figure \ref{fig:lowres_mslp}. Prior SOTA results~\cite{weyn2020improving} uses data at roughly this resolution to train their model, and we illustrate here that it is not possible to capture important small scale features such as hurricanes at this resolution. It is thus essential to train data driven models at a high resolution. 

\clearpage
\section*{Appendix C: ACC and RMSE definitions}
\customlabel{app:acc_rmse_define}{C}

The latitude weighted ACC for a forecast variable $v$ at forecast time-step $l$ is defined following~\citet{rasp2020weatherbench} as follows:

\begin{align}
\label{eq:acc}
\mathrm{ACC}(v, l)=\frac{\sum_{m, n} L(m) \tilde{\vv{X}}_{\text{pred}}(l)\left[ v, m, n \right] \tilde{\vv{X}}_{\text{true}}(l)\left[v, m, n\right]}{\sqrt{\sum_{m, n} L(m) \left( \tilde{\vv{X} }_{\text{pred}}(l)\left[v, m, n\right]\right)^{2} \sum_{m, n} L(m) \left(\tilde{\vv{X}}_{\text{true}}(l)\left[v, m, n\right]\right)^{ 2}}},
\end{align}

where $\tilde{\vv{X}}_{\text{pred/true}}(l)\left[v, m, n \right]$ represents the long-term-mean-subtracted value of predicted (/true) variable $v$ at the location denoted by the grid co-ordinates $(m, n)$ at the forecast time-step $l$. The long-term mean of a variable is simply the mean value of that variable over a large number of historical samples in the training dataset. The long-term mean-subtracted variables $\tilde{\vv{X}}_{\text{pred/true}}$ represent the anomalies of those variables that are not captured by the long term mean values. $L(m)$ is the latitude weighting factor at the co-ordinate $m$. The latitude weighting is defined by Equation~\ref{eq:latweight} as
\begin{align}
\label{eq:latweight}
L(j)=\frac{\cos (\operatorname{lat}(m))}{\frac{1}{N_{\text {lat }}} \sum_{j}^{N_{\text {lat }}} \cos (\operatorname{lat}(m))}.
\end{align} We report the mean ACC over all computed forecasts from different initial conditions and report the variability in the ACC over the different initial conditions by showing the first and third quartile value of the ACC in all the ACC plots that follow unless stated otherwise.

The latitude-weighted RMSE for a forecast variable $v$ at forecast time-step $l$ is defined by the following equation, with the same latitude weighting factor given by Equation~\ref{eq:latweight},
\begin{align}
\mathrm{RMSE}(v, l)= \sqrt{\frac{1}{NM} \sum_{m=1}^{M} \sum_{n=1}^{N} L(m)\left(\vv{X}_{\text{pred}}(l)[v,j,k]-\vv{X}_{\text{true}}(l)\left[v, j, k \right]\right)^{2}},
\end{align}

where  $\vv{X}_{\text{pred/true}}(l)\left[v, m, n \right]$ represents the value of predicted (/true) variable $v$ at the location denoted by the grid co-ordinates $(m, n)$ at the forecast time-step $l$.

\clearpage
\section*{Appendix D: Additional ACC and RMSE results}
\begin{figure}[h]
	\centering
	\includegraphics[width = \textwidth]{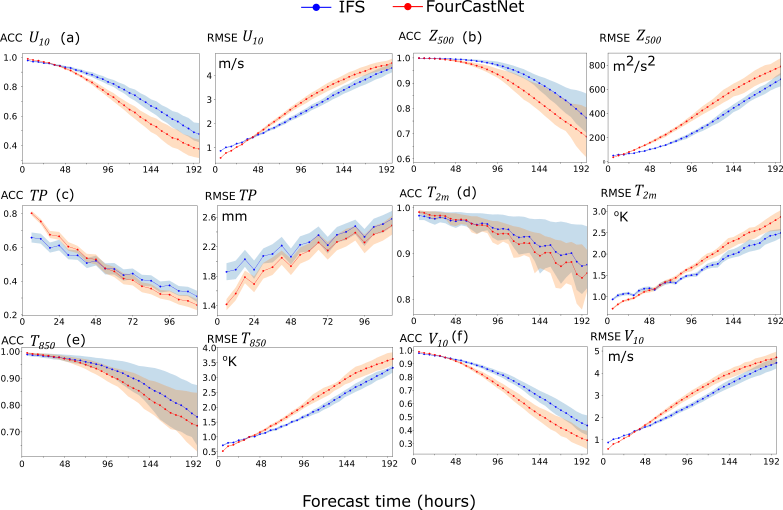}
	\caption{Latitude weighted ACC and RMSE curves for the FourCastNet model forecasts (Red line with markers) and the corresponding matched IFS forecasts (Blue line with markers) averaged over several forecasts initialized using initial conditions in the out-of-sample testing dataset corresponding to the calendar year 2018 for the variables (a) $U_{10}$, (b)  $TP$, (c) $T_{2m}$, (d) $Z_{500}$, (e) $T_{850}$, and (f) $V_{10}$. The ACC values are averaged over $N_f$ initial conditions with an interval of $D$ days between consecutive initial conditions. The $N_f$ and $D$ values are specified in Table~\ref{tab:IC}. The appropriately colored shaded regions around the ACC and RMSE curves indicate the region between the first and third quartile values of the corresponding quantity at each time step.}
	\label{fig:acc_rmse}
\end{figure}

\customlabel{app:acc_extra_results}{D}
Figures~\ref{fig:acc_rmse}(a-d) show the forecast skill of the FourCastNet model for a few key variables of interest along with the corresponding matched IFS forecast skill. Figure~\ref{fig:acc_rmse} is an extension of Figure~\ref{fig:IFS_AFNO} in the main text. In Figure~\ref{fig:acc_rmse}, we plot the latitude weighted RMSE and latitude-weighted ACC alongside each other for further clarity.

\begin{figure}[h]
    \centering
    \includegraphics[width = \textwidth]{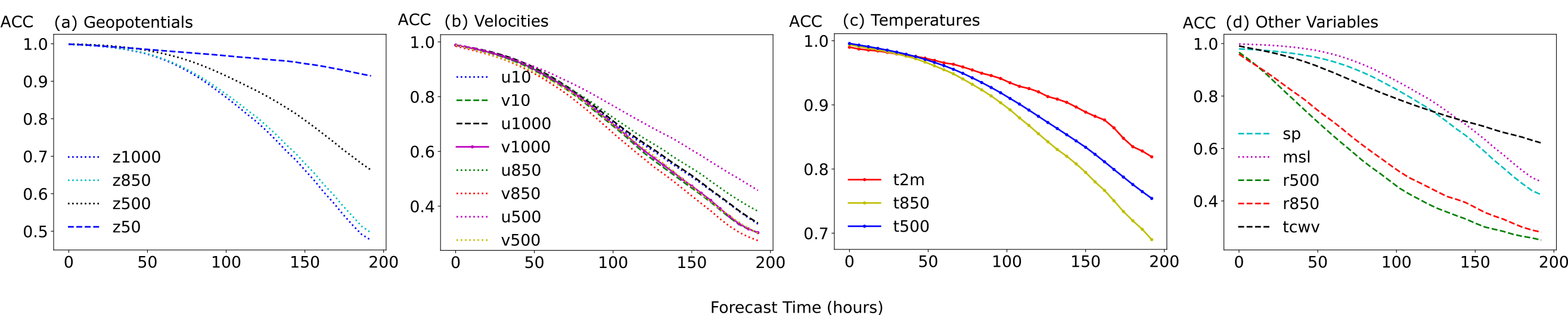}
    \caption{Panels (a)-(d) show the forecast skill of the FourCastNet model as measured by the latitude-weighted ACC for all the variables modeled by the backbone forecast model. Each ACC plot shows the mean ACC over $N_f =32$ initial conditions in the year 2018 with consecutive initial conditions spaced apart by an interval of $D = 9$ days. The The plots are grouped by similarity into (a) geopotentials, (b) wind velocities, (c) temperatures and (d) other variables. Panel (a) shows the ACC for wind velocities at the surface, 1000hPa level, 850hPa level and 500hPa level; Panel (b) shows the mean forecast ACC for the geopotentials at 1000hPa, 850hPa, 500hPa and 50hPa atmospheric levels; Panel (c) shows the mean ACC curves for the temperatures at the surface, 850hPa and 500hPa levels; Panel (d) shows the mean ACC curves for the surface pressure ($sp$), mean sea level pressure ($mslp$), relative humidity at 500hPa ($r500$) and 850hPa ($r850$), and the total column water vapor ($TCWV$).}
    \label{fig:acc_plots_othervars}
\end{figure}

Figures~\ref{fig:acc_plots_othervars}(a-d) show the forecast skill of the FourCastNet model for all the variables modeled by the backbone forecast model. The plots are grouped by similarity into wind velocities, geopotentials, temperatures and other variables. We see impressive performance across the variable set, with ACC generally staying above 0.6 for 5-10 days. Compared to the others, the relative humidity variables accumulate errors the fastest.

To assess our model accuracy over land and sea areas, we use the following equation for computing a land-specific and sea-specific ACC for surface wind velocity. 

\begin{align}
\label{eq:masked_acc}
\mathrm{ACC}_{\text{land/sea}}(v, l)=\frac{\sum_{m, n} \Phi^{m,n}_{\text{land/sea}} L(m) \tilde{\vv{X}}_{\text{pred}}(l)\left[ v, m, n \right] \tilde{\vv{X}}_{\text{true}}(l)\left[v, m, n\right]}{\sqrt{\sum_{m, n} \Phi^{m,n}_{\text{land/sea}} L(m) \left( \tilde{\vv{X} }_{\text{pred}}(l)\left[v, m, n\right]\right)^{2} \sum_{m, n} \Phi^{m,n}_{\text{land/sea}} L(m) \left(\tilde{\vv{X}}_{\text{true}}(l)\left[v, m, n\right]\right)^{ 2}}},
\end{align}

All the notation in Equation~\ref{eq:masked_acc} is the same as that in Eq~\ref{eq:acc} with the addition of a masking factor $\Phi^{m,n}_{\text{land/sea}}$. We use the land-sea mask provided in the ERA5 dataset as the land masking factor $\Phi^{m,n}_{\text{land}}$. The land masking factor is a static field with fraction of land in every grid box. The values are between 0 (grid box is fully covered with water) and 1 (grid box is fully covered with land). A grid box is considered to be land if more than 50\% of it is land, otherwise it's considered to be water (ocean or inland water, e.g. rivers, lakes, etc.). The corresponding sea masking factor $\Phi^{m,n}_{\text{sea}}$ is defined as $\Phi^{m,n}_{\text{sea}} = 1 - \Phi^{m,n}_{\text{land}}$.

\begin{figure}[h]
    \centering
    \includegraphics[width = 0.7\textwidth]{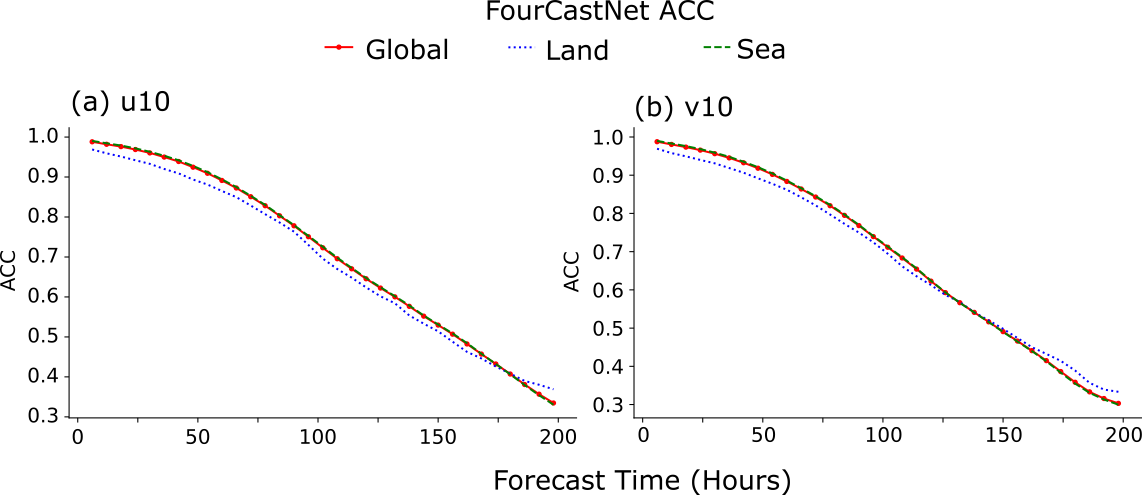}
    \caption{$\text{ACC}_{land}$, $\text{ACC}_{sea}$ as computed using Equation~\ref{eq:masked_acc} along with the overall ACC as computed using Equation~\ref{eq:acc} and averaged over $N_f = 32$ forecasts in the year $2018$ with consecutive forecast initial conditions separated by $D = 9$ days for (a) the meridonal velocity at 10m from the surface ($U_{10}$) and (b) the zonal velocity at 10 m from the surface ($V_{10}$).}
    \label{fig:acc_land_sea}
\end{figure}

In Figure \ref{fig:acc_land_sea}, we plot the $\text{ACC}_{\mathrm{land}}$ and $\text{ACC}_{\mathrm{sea}}$ for surface winds, and find that the FourCastNet model has very similar accuracy on forecasting ACC over landmass as it does over the ocean. This observation has significant implications for using the FourCastNet model in wind energy resource planning.

\end{document}